\documentclass[reprint,twocolumn,amsmath,amssymb,aps,prx]{revtex4-2}

\usepackage{xcolor}
\usepackage{graphicx}%
\usepackage{dcolumn}%
\usepackage{bibunits}
\usepackage{amsmath}
\usepackage{amssymb}
\usepackage{graphicx}

\usepackage{booktabs}

\usepackage{chngcntr}
\definecolor{NewBlue}{rgb}{0, 0, 0.41}
\definecolor{NewRed}{rgb}{0.6, 0.07, 0.07}

\usepackage{libertine}
\usepackage[colorlinks,
    linkcolor=NewBlue,
    citecolor=NewBlue,
    urlcolor=NewBlue]{hyperref}

\begin{document}
\newcommand{\C}{$^{13}\mathrm{C}\ $}
\newcommand{\TTwoStar}{$T_2^*\ $}

\title{Ten-second electron-spin coherence in isotopically engineered diamond} %

\author{T. Yamamoto$^{1,2}$}
\thanks{These authors contributed equally}
\author{H. B. van Ommen$^{1,2}$}
\thanks{These authors contributed equally}
\author{K.-N. Schymik$^{1,2}$}
\author{B. de Zoeten$^{1,2}$}
\author{S. Onoda$^3$}
\author{S. Saiki$^3$}
\author{T. Ohshima$^{3,4}$}
\author{H. Arjmandi-Tash$^5$}
\author{R. Vollmer$^5$}
\author{T. H. Taminiau$^{1,2,}$}
 \email{T.H.Taminiau@TUDelft.nl}

\affiliation{$^1$QuTech, Delft University of Technology, PO Box 5046, 2600 GA Delft, The Netherlands}

\affiliation{$^2$Kavli Institute of Nanoscience Delft, Delft University of Technology, P.O. Box 5046, 2600 GA Delft, The Netherlands}

\affiliation{$^3$Takasaki Institute for Advanced Quantum Science, National Institutes for Quantum Science and Technology (QST), Watanuki, Takasaki, Gunma 370-1292, Japan}

\affiliation{$^4$Department of Materials Science Tohoku University, Aoba, Sendai, Miyagi 980-8579, Japan}

\affiliation{$^5$Netherlands Organisation for Applied Scientific Research (TNO), P.O. Box 155, 2600 AD Delft, The Netherlands}

\date{\today}
\begin{abstract}
Solid-state spin defects are a promising platform for quantum networks. A key requirement is to combine long ground-state spin-coherence times with a coherent optical transition for spin-photon
entanglement. Here, we investigate the spin and optical coherence of single nitrogen-vacancy (NV) centres in (111)-grown isotopically engineered diamond. Our diamond-growth process yields a precisely controlled $^{13}$C concentration and low-ppb nitrogen concentrations. Combined with the mitigation of 50~Hz noise using a real-time feedforward scheme and tailored decoupling sequences, this enables record defect-electron-spin coherence times of $T_2 = 6.8(1)$~ms for a Hahn echo and of $T_2^{DD} = 11.2(8)$~s under dynamical decoupling. 
In addition, we observe coherent optical transitions with a near-lifetime-limited homogeneous linewidth of 16.9(4) MHz and characterize the spectral diffusion dynamics. These results provide new avenues to investigate the incorporation of impurities in diamond and new opportunities for improved spin-qubit control for quantum networks and other quantum technologies. 
\end{abstract}

\maketitle

\section*{Introduction}
\begin{figure*}
    \centering
    \includegraphics[width=15cm]{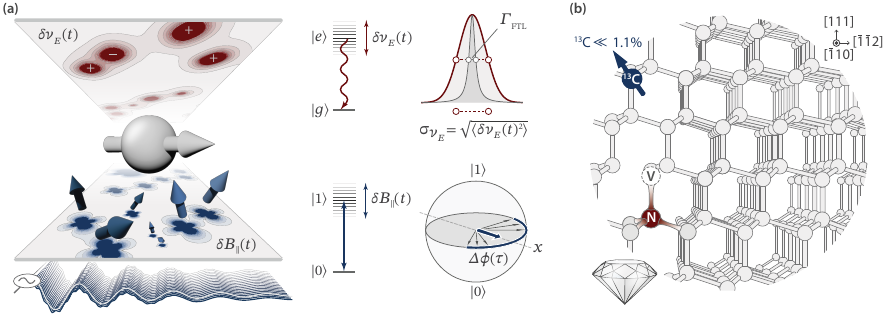}
    \caption{\emph{Noise and coherence for colour centres.} a) Colour centres are subject to electric field noise ($\delta \nu_E(t)$) and magnetic field noise ($\delta B_\parallel(t)$), which affect their optical and spin properties. Electric field noise primarily originates from charges moving around in the host material, and causes frequency broadening and diffusion of the optical transitions. Magnetic field noise is commonly caused by electron and nuclear spins in the host material, but external fields may also contribute, and is generally observed as decoherence of the ground-state spin qubit.
    b) In this work we investigate the optical and spin coherence of nitrogen vacancy (NV) centres in high-purity (111)-grown diamond with varying $^{13}$C nuclear spin concentrations. 
    }
    \label{fig:intro}
\end{figure*}

Solid-state colour centres provide a promising platform for quantum networks and distributed quantum computing \cite{wehnerQuantumInternetVision2018,knaut_entanglement_2024, pompili_realization_2021, abobeih_faulttolerant_2022,debone_thresholds_2024a}, as well as for quantum sensing and simulation \cite{lovchinsky_nuclear_2016, pieplowQuantumElectrometerTimeresolved2025, abobeih_atomicscale_2019, obergBottomupFabricationScalable2025}. A single colour centre constitutes a natural quantum network node, with an electron spin that forms a spin-photon interface \cite{stas_robust_2022,pompili_realization_2021, knaut_entanglement_2024} that is coupled to nuclear spins, which form a quantum register \cite{bradley_tenqubit_2019,abobeih_faulttolerant_2022, kunaLocalizationCoherentControl2025}. Recent breakthroughs include metropolitan-scale entanglement generation \cite{stolkMetropolitanscaleHeraldedEntanglement2024} and the demonstration of increasingly advanced quantum network protocols \cite{Weiuniversalblind2025}. A key requirement for a quantum network node is to combine long spin coherence for high-fidelity qubits \cite{debone_thresholds_2024a} with a coherent optical transition for spin-photon entanglement \cite{hermans_entangling_2023, babin_fabrication_2022, trusheim_transformlimited_2020, orphal-kobin_optically_2023,beukersRemoteEntanglementProtocolsStationary2024}, both of which benefit from realizing pure host materials with reduced sources of electric and magnetic noise (Fig. \ref{fig:intro}a).

Long electron-spin coherence times have been shown for in various platforms, including colour centres in diamond \cite{abobeih_onesecond_2018, bar-gillSolidstateElectronicSpin2013}, silicon-carbide \cite{andersonFivesecondCoherenceSingle2022}, and silicon \cite{bergeronSiliconIntegratedTelecommunicationsPhotonSpin2020}, and rare-earth ions \cite{ledantecTwentythreemillisecondElectronSpin2021}. Coherence times are typically extended by using pulse sequences, like dynamical decoupling, which protects from magnetic noise \cite{abobeih_onesecond_2018, andersonFivesecondCoherenceSingle2022}, and/or via isotopic purification of the host material to a spinless isotope, reducing the spin-bath noise \cite{bradley_robust_2022, herbschlebUltralongCoherenceTimes2019, andersonFivesecondCoherenceSingle2022,babin_fabrication_2022}. 
Despite this recent progress, it remains important to further understand, control and mitigate noise sources and prolong spin-coherence times, while preserving the colour centre's coherent optical transitions. 

Here, we investigate the optical and spin coherence of single nitrogen-vacancy (NV) centres in isotopically engineered, high-purity diamond. We demonstrate an electron-spin coherence up to 10 seconds, the longest reported for a solid-state electron-spin qubit, in combination with nearly lifetime-limited optical linewidths.

First, we present homoepitaxial diamond growth directly on (111) substrates, with precise control of the \C concentration (Sec. \ref{sec:growth}). $(111)$-oriented epilayers are crystallographically attractive for various types of quantum devices. However, achieving both high crystallinity and chemical purity is often considered more challenging in $(111)$ than in $(100)$ growth~\cite{Samlenski, PhysRevX.15.021035, Khmelnitsky}. Despite this, we find that our diamonds feature low strain and nitrogen concentrations in the low-ppb regime (Sec.~\ref{sec:Nestimation}).

Second, we study NV centres in these diamonds (Fig. \ref{fig:intro}b). We first examine the NV electron spin coherence for different \C concentrations (Sec. \ref{sec:spin}). We find that at low \C concentrations the Ramsey-coherence time $T_2^*$ and spin-echo time $T_2^{\mathrm{Hahn}}$ are limited by 50~Hz noise from the electrical grid. To mitigate this noise source we develop a feedforward compensation scheme, resulting in the longest Hahn-echo coherence time reported for colour centres ($T_2^{\mathrm{Hahn}} = 6.8(1)$~ms). While 50~Hz noise is commonly considered in, for example, atomic physics \cite{brandlCryogenicSetupTrapped2016, rusterLonglivedZeemanTrappedion2016}, these results highlight its --- perhaps underappreciated --- role in limiting colour-centre spin coherence. Using tailored decoupling sequences we extend the coherence time up to 11.2(8)~s. Then, we further confirm the high material quality by showing near-lifetime-limited optical linewidths (Sec. \ref{sec:optical}). We probe the remaining spectral diffusion, which we show is predominantly laser induced, and develop a model that gives insight into the timescales at which spectral diffusion becomes significant.
 
The realized combination of isotopic engineering and record-long coherence times, while maintaining optical coherence, provides new opportunities towards high-fidelity quantum network nodes and towards precise sensing of signals originating from in and outside the diamond.

\section{Isotopic control in $(111)$-oriented homoepitaxial diamond}
\label{sec:growth}
\begin{figure*}
  \centering
  \includegraphics[width=15cm]{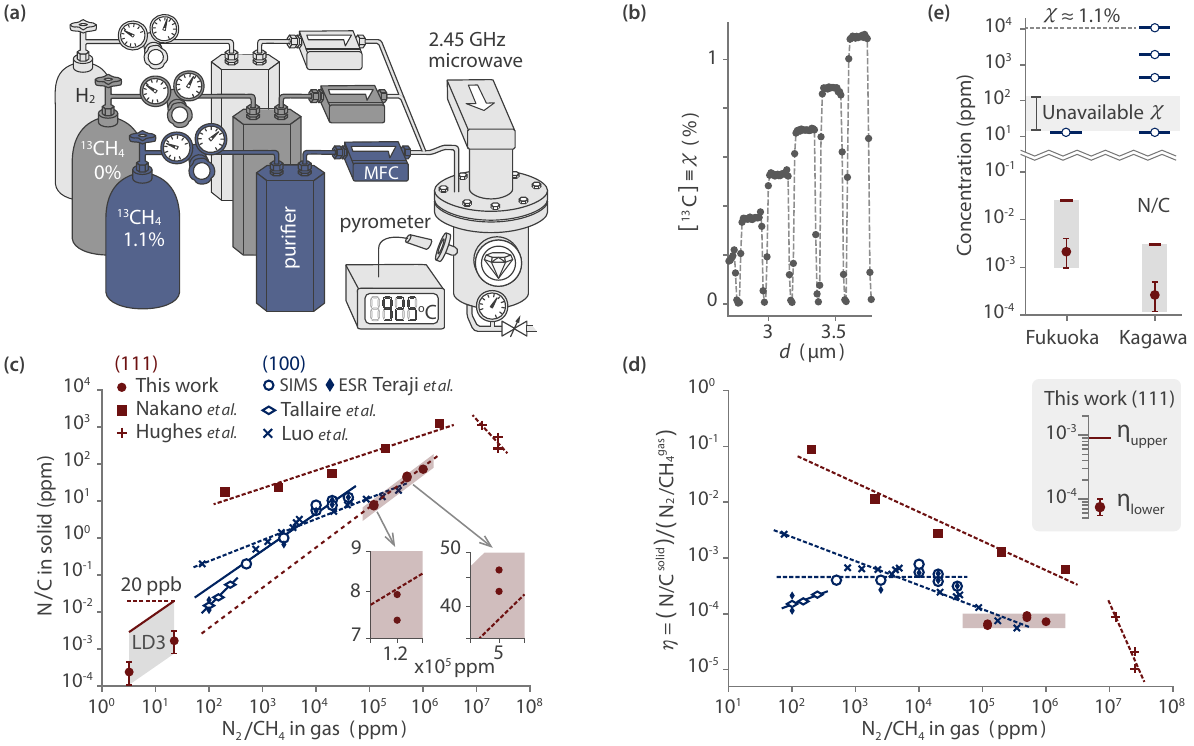}
  \caption{
\emph{Isotopically controlled diamond growth.}
(a) Schematic of the MPCVD system. The ratio of natural-abundance and $^{12}$C-enriched methane flow, set by the mass-flow controllers (MFCs), controls the diamond $^{13}$C concentration $\chi$. 
(b) SIMS depth profile of the $^{13}$C calibration sample LD3 with multiple $^{13}$C-coded layers.
(c) Solid-state N/C versus gas-phase N$_2$/CH$_4$, compiled from literature data and intentionally N-doped samples from this work.
For the calibration sample LD3, the horizontal dashed line indicates the SIMS detection limit, while the shaded region indicates the leak-limited N/C range estimated from the $\dot n_{\mathrm{N_2}}/\dot n_{\mathrm{CH_4}}$ ratio and the lower and upper bounds of the N-incorporation efficiency $\eta$.
Insets show magnified views of the intentionally N-doped samples; light red shading denotes the known day-to-day SIMS systematic uncertainty.
(d) $\eta = \mathrm{(N/C)_{\rm solid}/(N_2/CH_4)_{\rm gas}}$ extracted from the same dataset. The red shaded region indicates the range of $\eta_\mathrm{lower}$ estimated from the intentionally N-doped samples. The inset summarizes the bounds of $\eta$ for this work.
(e) Summary of the inferred N and $^{13}$C concentrations for the two NV samples studied (\textit{Fukuoka} and \textit{Kagawa}).
Blue markers indicate $\chi$ of each sample (Kagawa has 4 different layers). A small range of $\chi$ is inaccessible because of the MFC tuning ranges. Red dots (lower bounds) and lines (upper bounds) indicate the estimated N concentration. Error bars indicate the propagated 95\% confidence interval from both $\eta_{\rm lower}$ (from the intentionally N-doped samples) and the leak-derived $\dot n_{\mathrm{N_2}}$.
}
  \label{fig:growth}
\end{figure*}

To engineer a high-quality host material for single NV centres, we employ Microwave-Plasma Chemical Vapour Deposition (MPCVD~\cite{KAMO1983642}) diamond growth with a controlled \C concentration (Fig. \ref{fig:growth}a)~\cite{Watanabe}. Precise control of the \C concentration is an enabling capability for quantum network experiments, due to the ability to trade longer electron-spin and nuclear-spin coherence times with the abundance of $^{13}$C spins for use as qubits~\cite{bradley_robust_2022}. 

We chose to develop $(111)$-oriented diamond growth. While epitaxial growth on $(111)$ surfaces is more challenging than $(100)$ growth because of a higher tendency toward increased impurity incorporation \cite{Samlenski, PhysRevX.15.021035} and a narrower growth window for suppressing twinning-driven
morphological instabilities under high crystallization driving forces~\cite{Khmelnitsky}, it offers several advantages. These include improved optical detection into single-mode fibres~\cite{Aharonovich}, easier external magnetic field alignment with the $[111]$ NV axis, and preferential alignment of NV centres that are formed during CVD growth along the surface-normal $[111]$ direction~\cite{michlPerfectAlignmentPreferential2014, Lesik2014NV111alighment, fukuiPerfectSelectiveAlignment2014, miyazakiAtomisticMechanismPerfect2014}. Compared to cleaving $(100)$-grown diamonds, direct $(111)$-oriented growth enables large isotopically engineered layers and precise depth control of isotopes.

We grow homoepitaxial diamond layers on low-dislocation $(111)$ single-crystal diamond substrates (see Methods for details).
This approach leverages the intrinsically high crystalline quality of high-pressure high-temperature (HPHT) diamond while enabling isotope engineering through MPCVD overgrowth.
A custom mounting stage was employed to maintain a similar plasma--sample distance and geometry across growth runs, thereby minimizing run-to-run variations in the near-surface plasma chemistry.
We opted for growth under low methane concentrations (CH$_4$/H$_2 < 0.08\%$) resulting in a relatively slow growth rate ($<0.7~$\textmu m/h) and favouring step-flow growth while suppressing secondary nucleation on the $(111)$ surface ~\cite{TOKUDA2010288}.
This growth mode, combined with the use of low-dislocation substrates, yields smooth epilayers with a low overall dislocation density~\cite{FRIEL2009808}. Such epilayers can reduce impurity trapping at dislocation cores~\cite{PhysRevB.106.174111} and suppress morphological instabilities that influence impurity incorporation~\cite{10.1063/5.0230723}.

The $^{13}$C concentration in the epilayer was controlled by mixing natural-abundance and $^{12}$C-enriched methane, each supplied through an independent mass flow controller (MFC),
with flow-rate ranges of 0.06--3 and 0.1--5~sccm (standard cubic centimetres per minute), respectively
(Fig.~\ref{fig:growth}a). We grow a calibration sample (named LD3) with a number of layers grown at varying flow ratios $f_i = f_1/f_0$, where $f_0$ ($f_1$) denotes the setpoint flow rate of $^{12}$C-enriched (natural-abundance) methane. The $^{13}$C isotope concentration is subsequently analysed using Secondary Ion Mass Spectrometry (SIMS) (Fig. \ref{fig:growth}b). The raw isotope intensity measured with the SIMS machine is calibrated to a reference sample at natural abundance to express the measured $^{13}$C isotope concentration relative to Vienna PeeDee Belemnite (VPDB), a commonly used reference for the $^{13}\mathrm{C}/^{12}\mathrm{C}$ ratio~\cite{VPDB} (see methods for the calibration procedure).

We define the $^{13}$C concentration as $\chi={}^{13}\mathrm{C}/({}^{13}\mathrm{C} + {}^{12}\mathrm{C})$. Using only natural-abundance methane yields $\chi_1 = 1.0937(86)\%$, while only using $^{12}$C-enriched methane yields $\chi_0=0.0013(3)\%=13(3)$ ppm. When mixing the gases, the measured ${\chi_i}$ (Fig. \ref{fig:growth}b) is reproduced by introducing an effective flow ratio $\tilde f_i = f_1 / f_0'$ such that
\begin{equation}
  \chi_i = \frac{\chi_0 + \tilde f_i\,\chi_1}{1 + \tilde f_i},
\qquad
  f_0' = 1.023\, f_0 + 0.036.
\end{equation}
With this definition the accessible $^{13}$C tuning range is 0.0139--1.0456\%, which is constrained by the flow rate setting ranges of the MFCs.

\section{Nitrogen concentration}
\label{sec:Nestimation}
SIMS measurements of impurities other than $^{13}\mathrm{C}$ fall below the instrumental background ($<200$~ppb H, $<20$~ppb N, $<10$~ppb B). While the concentrations are low, the presence of these impurities may nevertheless be relevant for the spin or charge dynamics in the diamond, as well as for the local strain environment.
Nitrogen is a particularly important contaminant due to its prevalence in air which inevitably leaks into the CVD reactor, and because it creates an electron-spin bath and a fluctuating charge environment that affect the spin and optical coherence of defect centres \cite{wangSpinDecoherenceElectron2013, orphal-kobin_optically_2023}. Therefore a more precise determination of the nitrogen concentration is desirable. Although sub-ppb nitrogen concentrations can be detected by ESR in bulk diamond samples of tens of milligrams~\cite{PhysRevMaterials.2.094601}, the present samples, with approximately two orders of magnitude smaller mass, are expected to be sensitivity-limited for direct ESR measurements. In the following we place an upper and lower bound on the nitrogen concentration by estimating the amount of nitrogen in the process gas, and the efficiency of incorporation into the diamond.

We use the nitrogen incorporation efficiency $\eta$ to relate the gas-phase N$_2$/CH$_4$ ratio during growth to the solid-state nitrogen concentration in the diamond epilayer after growth (equivalently expressed as an N/C ratio)\begin{equation}
    [\mathrm{N}]_{\rm ppb}
= \eta
  \frac{\dot n_{\rm N_2}}{\dot n_{\rm CH_4}}
  \times 10^9,
\end{equation}
where $\dot n_{\rm CH_4}$ ($\dot n_{\rm N_2}$) is the methane (nitrogen) molar flow rate during growth. $\dot n_{\rm CH_4}$ is set by the MFC-regulated flow rate. We estimate $\dot n_{\rm N_2}$ by considering the vacuum leak rate of the reactor chamber, outgassing from chamber materials, and impurity of the CH$_4$ and H$_2$ source gases (\ref{sec:app_nitrogen_outgas}). We find that air leaking into the reactor is the dominant source of nitrogen, with a nitrogen inflow rate of \begin{equation}\label{eq:nitrogen_flow}
    \dot n_{\rm N_2} = 4.0(1.6)\times10^{-12}~\mathrm{mol}\,\mathrm{s}^{-1}.
\end{equation}
Without purposeful nitrogen doping, $[\mathrm{N}]$ is therefore determined by the chosen methane flow rate $\dot n_{\rm CH_4}$ and the incorporation efficiency $\eta$.

We use two methods to estimate $\eta$, thereby placing lower and upper bounds on $[\mathrm{N}]$. A lower bound on $\eta$ is obtained from SIMS measurements of intentionally N-doped samples. An upper bound on $\eta$ is derived from the $^{13}$C calibration sample, which was not intentionally N-doped, based on the SIMS constraint of $[\mathrm{N}] < 20\,\mathrm{ppb}$.

Figures~\ref{fig:growth}c,d summarize our measurements of $[\mathrm{N}]$ and the extracted $\eta$ together with literature results. The literature data include $[\mathrm{N}]$ extracted from SIMS~\cite{PhysRevX.15.021035,Nakano,Teraji} and ESR~\cite{Teraji,TALLAIRE20061700,Luo_2022} for both (111) and (100) growth orientations (note that ESR only detects $P_1$ centres, $\sim75\%$ of $[\mathrm{N}]$ \cite{Luo_2022, Teraji}). The dashed lines serve as guides to the eye, indicating overall trends in the log--log representation.

The lower-bound $\eta$ estimation is based on three N-doped samples (solid red circles; see Methods for growth details). Two were grown under identical N$_2$/CH$_4$ conditions, each with two doped layers at $1.2\times10^5$ and $5\times10^5$~ppm in the gas phase, while the third had a single layer at $10\times10^5$~ppm. The measured N concentration follows a power-law, $[\mathrm{N}] \propto (\mathrm{N}_2/\mathrm{CH}_4)^{1.08}$, similar to the $\propto (\mathrm{N}_2/\mathrm{CH}_4)^{1.00}$ dependence reported by Teraji \emph{et al.} for $(100)$ growth~\cite{Teraji}, albeit with a reduced prefactor. This scaling implies a constant incorporation efficiency \begin{equation}
\eta = 7.5^{+2.6}_{-2.0}\times10^{-5}
\end{equation}
(95\% confidence intervals indicated), which we treat as a lower bound, because $\eta$ may increase for lower N$_2$/CH$_4$ ratios where we have no available data \cite{VincentMortet}.

The upper-bound on $\eta$ is obtained from the $[\mathrm{N}]<20$~ppb SIMS background limit measured on the $^{13}$C calibration sample. Because the CH$_4$ flow was varied across the \C stack ($\dot n_{\mathrm{CH_4}}=0.24-1.67$~sccm) while $\dot n_{\mathrm{N_2}}$ is constant, the inferred gas-phase $\dot n_{\mathrm{N_2}}/\dot n_{\mathrm{CH_4}}$ ratio also varies between the layers. The lowest $\dot n_{\mathrm{CH_4}}$ combined with the vacuum leak rate $\dot n_{\mathrm{N_2}}$ means the maximum $\dot n_{\mathrm{N_2}}/\dot n_{\mathrm{CH_4}} \simeq 23$~ppm. Combined with the SIMS background limit, this results in an upper limit of $\eta < 8.9\times10^{-4}$. Here we assume that $\eta$ \textit{remains} constant over the inferred N$_2$/CH$_4$ range explored in this work for the non-intentionally N-doped growth conditions.

Taken together, $\eta$ is constrained to  
\begin{equation}
0.55\times10^{-4} < \eta < 8.9\times10^{-4},
\end{equation}
which is lower than, or comparable to values reported for (100) growth. Using this $\eta$ we can estimate $[\mathrm{N}]$ in the $^{13}$C calibration sample (LD3 in Fig.~\ref{fig:growth}c), which is representative of our growth process. The range of used $\dot n_{\mathrm{CH_4}}$ flow rates results in a range of $[\mathrm{N}]$.
The lower bound on $\eta$ implies $[\mathrm{N}] = 0.2\text{--}2.3~\mathrm{ppb}$, while the upper bound gives 2.9--20~ppb. For comparison, specifications of commercial $\{100\}$ electronic-grade diamond are typically below 5~ppb [N] (Element Six), and ESR measurements on one such sample have reported P1 concentrations as low as $0.07\pm0.02$~ppb~\cite{PhysRevMaterials.2.094601}.%

Although the theoretical nitrogen-incorporation pathways differ substantially between (111) \cite{miyazakiAtomisticMechanismPerfect2014} and (100) \cite{Kelly2017,OBERG2021606} growth orientations, the resulting $[\mathrm{N}]$ is likely governed primarily by the local ratio of nitrogen- to carbon-containing reactive species at the growth surface. This interpretation is consistent with both our compilation of reported data in Fig.~\ref{fig:growth}(c,d) and a broader literature survey \cite{VincentMortet}, which show no clear separation in $[\mathrm{N}]$ between (111) and (100) growth. The low $[\mathrm{N}]$ obtained here likely reflects a plasma--surface environment that is less favourable for N incorporation relative to carbon growth in comparison to other reported (111) growth conditions. Because this local N-to-C reactive-species ratio (represented here by atomic N and CH$_3$, respectively \cite{Truscott2016}) can vary spatially \cite{ASHFOLD2023110097}, $[\mathrm{N}]$ is also likely to depend sensitively on the plasma--surface distance and sample-holder geometry, even under otherwise similar growth conditions. Finally, increasing the substrate temperature may further reduce $[\mathrm{N}]$, even in (111) growth \cite{YAMADA201327}.

\section{NV-centre Samples}
Next, we investigate the spin and optical properties of NV centres in two diamond samples (Fig.~\ref{fig:growth}e).

Sample \textit{Fukuoka} was grown at the lowest possible $^{13}$C concentration $\chi=0.0013(3)\%$. Based on the CH$_4$ flow rate (0.19~sccm) the lower-bound $\eta$ gives an estimated lower-bound $[\mathrm{N}]$ of 1--4~ppb, while our upper-bound estimate is $[\mathrm{N}]<26$~ppb. Single NV centres are present in the diamond as-grown and are preferentially aligned along the surface-normal $[111]$ direction (confirmed by single-NV ESR measurements), as is characteristic of $(111)$ growth \cite{michlPerfectAlignmentPreferential2014, Lesik2014NV111alighment, fukuiPerfectSelectiveAlignment2014, miyazakiAtomisticMechanismPerfect2014}. 
Solid-Immersion Lenses (SILs) were fabricated to enhance the collection efficiency \cite{hadden_strongly_2010} and striplines were fabricated for MW control of the NV centre electron spin (see methods for fabrication details).

Sample \textit{Kagawa} features four different layers with $\chi=[0.0442, 1.0937, 0.1949,0.0013]\%$. A higher CH$_4$ flow rate
(1.53~sccm) results in a lower inferred $[\mathrm{N}]$, with a lower-bound estimate of 0.1--0.5~ppb, and an upper-bound estimate of $[\mathrm{N}]<3.2$~ppb.
No naturally occurring NV centres were detected in the sample, consistent with the lower $[\mathrm{N}]$ estimate than \textit{Fukuoka}. Instead, NV centres were created by 2\,MeV electron irradiation (fluence $5\times10^{11}$\,cm$^{-2}$), followed by annealing at 1000$^\circ$C for 2\,h in vacuum.

\section{Electron-spin Coherence}
\label{sec:spin}
\begin{figure*}
    \centering
    \includegraphics{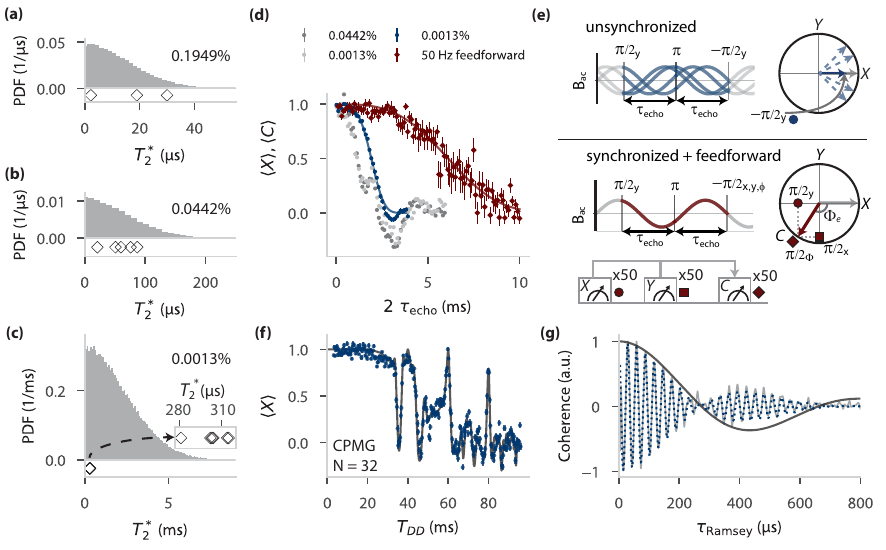}
    \caption{\emph{Spin coherence in isotopically purified diamond.} a,b,c) Expected $T_2^*$ distribution (grey histogram) and measured $T_2^*$ times (diamonds) for NV centres at $\chi=0.1949\%$ (a, sample Kagawa, 3 measurements), $\chi=0.0442\%$ (b, Kagawa, 5 measurements), and $\chi=0.0013\%$ (c, Fukuoka, 6 measurements). At the highest degree of purification, $T_2^*$ is limited to roughly $300\,$ \textmu s. An inset is used in c to more clearly show the $T_2^*$ times. d) Hahn-echo data showing $T_2^\mathrm{Hahn} \approx 2$~ms in Kagawa (grey) and in Fukuoka in a SIL NV (blue). We attribute the difference in data to the use of a different measurement setup. (e) When the Hahn-echo sequence in \textit{unsynchronized}, external ac-fields cause dephasing. By triggering the start of the echo sequence with a clock that is \textit{synchronized} to the mains frequency ($\sim50$~Hz), interference from the mains-frequency ac-fields can be measured and compensated using a feedforward scheme. We measure $\left<X\right>$ and $\left<Y\right>$ with 50 shots to estimate $\Phi_e$, and then measure the corrected Pauli observable $\left<C\right>$ along the axis defined by $\Phi_e$. This prolongs $T_2^\mathrm{Hahn}$ up to $6.8(1)$~ms (red diamonds in d). Errorbars are derived from the standard deviation over 12 measurements of 50 shots each. (f) DD spectroscopy probing the mains interference. A model assuming 50~Hz ac fields (solid black line) was fit to the data, matching the measurements well. Note the revivals for 20, 40, 60 and 80~ms. (g) Ramsey signal of a SIL NV at $\chi=0.0013\%$. We compare the data (grey line) to a fit using a product of two cosines and gaussian exponential decay (dashed blue line, $T_2^*=383.4(7)$~\textmu s), and to a prediction of the coherence envelope (dark grey line) by the 50~Hz noise model.}%
    \label{fig:spin}
\end{figure*}
We investigate the electron-spin coherence of single NV centres and probe the noise sources that limit the Ramsey (\TTwoStar), Hahn-echo ($T_2^\mathrm{Hahn}$), and CPMG dynamical decoupling (DD, $T_2^\mathrm{DD}$) coherence times. When the \C spin bath is the limiting cause of dephasing, the coherence times are expected to scale with the \C concentration $\chi$ according to $T_2^*\propto\chi^{-1}$ \cite{zhao_decoherence_2012} and $T_2^\mathrm{Hahn}\propto\chi^{-1}$ \cite{ShunKanai2022pnas}.

First, the electron-spin \TTwoStar of NV centres at three different $\chi$ was measured (Fig. \ref{fig:spin}a-c). For a given $\chi$, \TTwoStar for different NV centres will follow a probability distribution due to the random configuration of the nuclear spin environment of each NV centre (assuming that no other noise sources are present). We simulate the expected distribution (details in \ref{sec:app_t2star_sim}). \TTwoStar follows the simulated distribution well for $\chi=0.1949\%$ and $\chi=0.0442\%$, but for the lowest concentration $\chi=0.0013\%$, the measurements are clustered around $T_2^* \approx 300$~\textmu s. The clustering, and \TTwoStar being shorter than expected ($\sim$ 2~ms), together indicate that another noise source limits \TTwoStar. We rule out that an electron-spin bath, arising from $P_1$ centres or vacancy complexes, limits \TTwoStar, as it cannot explain the clustering \cite{marcksGuidingDiamondSpin2024}. Nevertheless the measurements of \TTwoStar place an upper bound of $\sim21$~ppb on the electron-spin bath concentration (\ref{sec:app_t2star_espin}).

Hahn-echo measurements showed $T_2^\mathrm{Hahn}$ times that were limited to $\sim2$~ms (Fig. \ref{fig:spin}d), which is shorter than expected from just the \C bath (up to $\sim800$~ms for $\chi=0.0013\%$ \cite{ShunKanai2022pnas}). Other studies on defect centres in multiple different isotopically purified materials have reported similar $T_2^\mathrm{Hahn}$ times (in the 1.5--2.5~ms range) \cite{bar-gillSolidstateElectronicSpin2013, balasubramanianUltralongSpinCoherence2009, ishikawaOpticalSpinCoherence2012, herbschlebUltralongCoherenceTimes2019, andersonFivesecondCoherenceSingle2022, bergeronSiliconIntegratedTelecommunicationsPhotonSpin2020}. We next show that magnetic fields synchronous to the $\sim50$~Hz frequency of the mains electricity grid are limiting our observed $T_2^\mathrm{Hahn}$ and \TTwoStar. We note that, although the effect of mains-frequency signals on qubit coherence is commonly considered in fields such as atomic physics \cite{brandlCryogenicSetupTrapped2016, rusterLonglivedZeemanTrappedion2016}, the recurring observation of limited $T_2^\mathrm{Hahn}$ in the literature suggest that this might have been overlooked for colour centres. While one mitigation path is to physically reduce the 50~Hz interference, e.g. through shielding or moving equipment, here we choose instead to mitigate this noise through a qubit-based feed-forward scheme and tailored pulse sequences.

The mains interference can be described by an ac magnetic field along the z-axis of the NV
\begin{equation}
    B_{ac}(t) = \sum_{i=1}^n B_i \cos(\omega_i(t-t_0) +\phi_i),
\end{equation}
where the summation is performed over $n$ frequency components at 50~Hz and higher harmonics ($\omega_i$), each with amplitude $B_i$ and relative phase $\phi_i$. $t_0$ is the delay between the start of the pulse sequence ($t=0$) and a reference point in the mains waveform. In a spin echo, the electron spin picks up a phase $\Phi_e$ according to \begin{equation}
    \Phi_e = \gamma_{\mathrm{NV}}\left[\int_0^\tau B_{ac}(t) \mathrm{d}t  - \int_\tau^{2\tau} B_{ac}(t) \mathrm{d}t \right],
\end{equation}
where $\gamma_{\mathrm{NV}}$ is the NV centre gyromagnetic ratio, and $\tau$ is the echo time. The spin-echo measures $\left<X\right> = \cos \Phi_e$. 

The fact that 50~Hz interference is limiting the coherence can be seen by synchronizing the start of the spin echo to the mains frequency (Fig. \ref{fig:spin}e). When the sequence is unsynchronized, $t_0$ is randomized every measurement repetition, causing a Bessel-like dephasing for $\left<X\right>$. When instead the sequence is synchronized, here achieved by triggering the spin-echo sequence at a positive zero-crossing of the 50~Hz mains waveform, $t_0$ is fixed, causing $\Phi_e$ to be deterministic. $\left<X\right>$ then oscillates coherently as a function of $\tau$, assuming that $B_i$ and $\phi_i$ are constant over the course of the measurement.

The synchronization allows a feedforward scheme that partly corrects the 50~Hz interference. For a specific spin-echo time $\tau$, we measure $\left<X\right>$ and $\left<Y\right>$ to estimate $\Phi_e$, with 
\begin{equation}
    \Phi_e = \arctan{\frac{\left<Y\right>}{\left<X\right>}}.
\end{equation}
The 50~Hz interference is then corrected during subsequent echo experiments by shifting the phase of the final $\pi/2$-pulse by $-\Phi_e$, yielding a new expectation value for the spin coherence $\left<C\right>$.

For a SIL NV at $\chi=0.0013\%$, $\left<X\right>$, $\left<Y\right>$, and $\left<C\right>$ are each measured with 50 shots (Fig. \ref{fig:spin}d, red diamonds). The feed-forward $\left<C\right>$ data results in $T_2^\mathrm{Hahn} = 6.8(1)$~ms, the longest reported for a defect centre electron-spin qubit. The remaining decay and larger spread in $\left<C\right>$ are caused by fluctuations in the ac-field magnitudes $B_i$ during the $\Phi_e$ estimation, which occur on the time scale of a few seconds (\ref{sec:app_echoData}). The choice for the number of shots (20~ms each due to the 50~Hz synchronization) is a trade-off between the accuracy with which $\Phi_e$ is determined and the time-scale of fluctuations in $B_i$ that can be tracked.

\begin{figure*}
    \centering
    \includegraphics{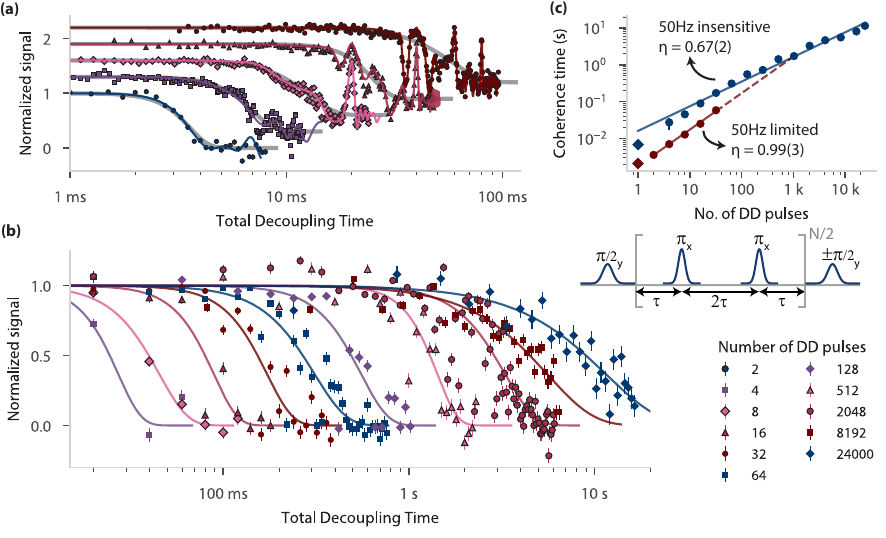}
    \caption{\emph{Dynamical decoupling.} a) High resolution CPMG data at $\chi = 0.0013\%$, showing the collapses and revivals due to 50~Hz noise. Data are each offset by 0.3 on the y-axis for clarity. Coloured solid lines are predictions from the 50~Hz model ($N=32$ data also shown in Fig. \ref{fig:spin}f). The grey curves are decay fits of the data (Eq. \ref{S_decay}) when the signal revivals are ignored. b) CPMG data with more decoupling pulses. Up to $N=2048$, 50~Hz noise was mitigated by choosing interpulse delays $\tau$ such that $T_\mathrm{DD}$ is a multiple of 20~ms. Data and fits are rescaled by the fit parameter $A$. Inset: Dynamical decoupling sequence. c) Extracted coherence time versus the number of decoupling pulses. The coherence time follows $T_2^\mathrm{DD}=T_0 N^\eta$. In the case where 50~Hz noise is not mitigated by taking data at revival times (grey lines in a), $T_2^\mathrm{DD}$ scales linearly with $N$ (red line, $T_0=1.8(1)$~ms). Mitigating 50~Hz noise results in $\eta=0.67(2)$ (fits from b, solid blue line, $T_0=16(2)$~ms). Diamond data points at $N=1$ are the $T_2^\mathrm{Hahn}$ times with (blue) and without (red) feedforward correction for 50~Hz noise.}
    \label{fig:dd}
\end{figure*}

We then performed CPMG dynamical decoupling to further extend the coherence time and probe the noise spectrum (Fig. \ref{fig:spin}f and Fig. \ref{fig:dd}a). These measurements were not synchronized to the mains frequency. The phase $\Phi_\mathrm{DD}$ picked up by the spin can be described by $B_{ac}(t)$ and the filter function of the CPMG sequence $F(\omega, N, \tau)$, with $N$ the number of $\pi$-pulses, and $2\tau$ the delay between $\pi$ pulses (see \ref{sec:app_50hz} for details). This enables a fine-grained analysis of $B_{ac}(t)$. We fit a model of the mains interference to the CPMG data (up to $N=32$), assuming that $B_{ac}(t)$ is described by 50~Hz harmonics up to 450~Hz, with $B_i$ and $\phi_i$ for each harmonic as a free parameter (Table \ref{tab:50model}). The model reproduces the data well, confirming that 50~Hz interference is the dominant noise source for CPMG measurements at these values for $N$.

The $B_{ac}(t)$ model allows us to consider the effect of the 50~Hz interference on Ramsey measurements. We observe a decaying beating signal (Fig. \ref{fig:spin}g).
A naive fit to a product of two cosines yields a decay time of $T_2^*=383.7(7)$ \textmu s. However, a comparison to the predicted signal envelope, based on $B_{ac}(t)$ obtained from the CPMG data (Table \ref{tab:50model}), indicates that the 50~Hz interference also is the external noise source that limits the Ramsey signal and $T_2^*$ (Fig. \ref{fig:spin}c). 
Note that the predicted envelope is averaged over a range of amplitude fluctuations of $B_{ac}(t)$, as observed in the Hahn-echo data (\ref{sec:app_echoData}).

\begin{table}
    \centering
    \begin{tabular}{llll}\toprule
         Frequency (Hz)&   $B_{ac}$ (mG) &$\gamma_{\mathrm{NV}} B_{ac}$ (kHz)&$\phi_{ac}$ (rad)\\\midrule
         50&  2.95(3) &8.28(8)&0\\
 100& 0.024 &0.067&3(5)\\
 150& 0.490(6) &1.37(2)&-1.77(3)\\
 200& 0.0065(8) & 0.018(2) & -0.4(8)\\
 250& 0.046(2) &0.129(4) &4.33(9)\\
 300& 0.010(2) & 0.029(6) &0(3)\\
 350& 0.0376(1) &0.105(3)&0.9(3)\\
 450& 0.0409(1) &0.115(3)&-1.3(4)\\ \bottomrule \end{tabular}
    \caption{\emph{Parameters for the 50~Hz interference model.} The phase of the 50~Hz component is chosen to be 0. Both the extracted field amplitude $B_{ac}$, and the corresponding shift on the electron-spin frequency $\gamma_\mathrm{NV}B_{ac}$ are reported. The amplitude of the 100Hz component is chosen to match the CPMG data at $\tau=2.5\, \mathrm{ms}$ (\ref{sec:app_50hz}), where only 100~Hz noise is detected.
    }
    \label{tab:50model}
\end{table}

Finally, we investigate the coherence with an increasingly higher number of pulses $N$ (Fig.~\ref{fig:dd}b, complete dataset in \ref{sec:dd_complete}). To avoid the 50~Hz noise, we set the interpulse delays $\tau$ to follow the revivals at $T_\mathrm{DD}=2N\tau = k\times 20$~ms ($k \in \mathbf{N}$) (Fig.~\ref{fig:dd}a). These revivals occur because, for most $\tau$ at these $T_\mathrm{DD}$, the CPMG filter function $F(\omega, N, \tau)=0$ for $50$~Hz harmonics (See \ref{sec:app_50hz} for exact conditions).
We fit the function \begin{equation}\label{S_decay}
    S(T_\mathrm{DD})=A \exp\left[-(T_\mathrm{DD}/T_2^\mathrm{DD})^n\right]
\end{equation} to the data, with $A$, $T_2^\mathrm{DD}$, and $n$ free parameters. %
Increasing the number of decoupling pulses to 24000 results in a coherence time of 11.2(8)~s, the longest coherence time reported for a single electron spin in a solid \cite{abobeih_onesecond_2018,andersonFivesecondCoherenceSingle2022}. 

Further insight can be obtained by studying the scaling of the coherence time with the number of decoupling pulses $T_2^\mathrm{DD} = T_0 N^\eta$ (Fig. \ref{fig:dd}c). As a comparison, we first consider the hypothetical case in which the pulse sequence is not tailored to avoid 50~Hz noise, and a naive dephasing curve is fit to the initial coherence decay while ignoring (or missing) the sharp signal revivals (Fig. \ref{fig:dd}a, grey lines). The coherence time then scales linearly with pulse number $\eta\approx1$. We note that this is conceptually similar to initial collapse of the electron-spin coherence in a nuclear spin bath \cite{ryanRobustDecouplingTechniques2010}.  
Taking data at the revival times, we find $\eta = 0.67(2)$. Such a scaling is consistent with the $\eta=2/3$ scaling caused by a Lorentzian ($\omega^{-2}$) noise spectral density, as would be caused by e.g. an electron-spin bath \cite{desousaElectronSpinSpectrometer2009, delange_universal_2010, biercukPhenomenologicalStudyDecoherence2011}. The two observed scalings show that beyond $N\approx1000$ the 50~Hz interference is no longer the limiting noise source for the $T_2^{\mathrm{DD}}$, and thus it is no longer necessary to follow the signal revivals. 
Separate scaling behaviour for low and high $N$ is also reported in other work \cite{andersonFivesecondCoherenceSingle2022}; 50 (or 60)~Hz noise might have also played a role in those results (see \ref{sec:app_CPMGcomp}). Assuming that the 2/3-scaling corresponds to a $[\mathrm{N}]$-limited coherence regime we can extrapolate a hypothetical $T_2^{\mathrm{Hahn}} = 16$~ms from the fit. Using the (NV ensemble) results from Bauch \textit{et al.} \cite{bauchDecoherenceEnsemblesNitrogenvacancy2020} which relate $T_2^\mathrm{Hahn}$ and $[\mathrm{N}]$ yields $[\mathrm{N}]\approx 10$~ppb, which is in agreement with our estimated range for $[\mathrm{N}]$ (Fig. \ref{fig:growth}e).

\section{Optical Properties}
\label{sec:optical}
\begin{figure*}
    \centering
    \includegraphics{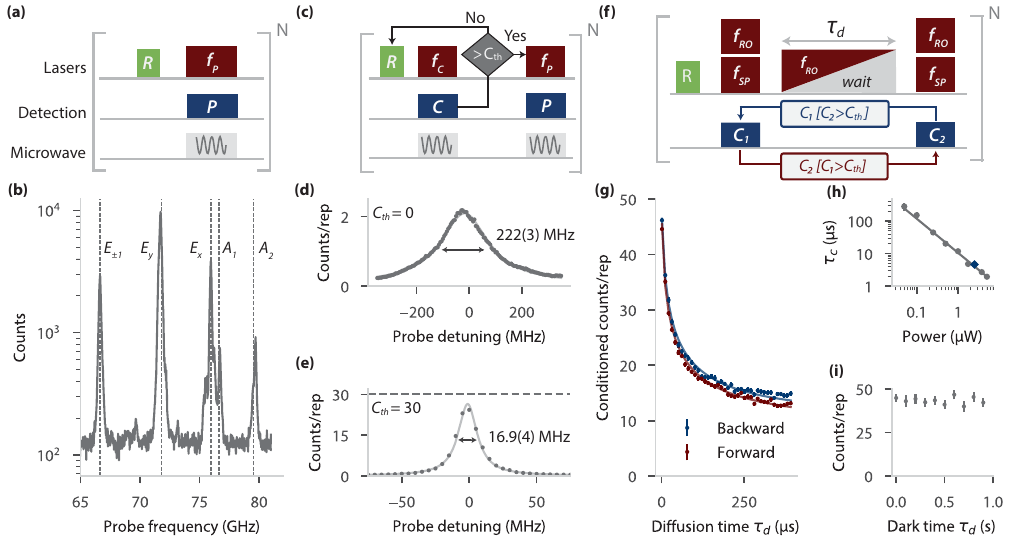}
    \caption{\emph{Optical coherence and spectral diffusion.} a) Diagram for a PLE scan. A repump laser pulse $R$ at 515~nm ($20\,$\textmu W, $50\,$\textmu s) is followed by a \textit{probe} pulse at frequency $f_p$, resulting in $P$ detected photons. Microwaves at the $m_s=\pm1\rightarrow m_s=0$ are applied during the probe pulse. b) PLE scan showing a low-strain NV optical spectrum. c) Check-Probe measurement scheme~\cite{vandestolpeCheckprobeSpectroscopyLifetimelimited2025}. $f_c$ is kept at the $E_y$ transition, while the detuning $f_c-f_p$ is varied using $f_p$ (both pulses $400\,$\textmu s at 1~nW). Each $f_p$ is measured $N=50000$ times. d) Check-probe PLE of $E_y$. Without thresholding ($C_{th}=0$), the inhomogeneous linewidth is $222(3)$ MHz (Lorentzian fit). e) Using $C_{th}=30$ gives a homogeneous linewidth of $16.9(4)$ MHz with a check success probability of 1.4\%. f) Spectral-diffusion measurement scheme~\cite{vandestolpeCheckprobeSpectroscopyLifetimelimited2025}. Check blocks before ($C_1$) and after ($C_2$) the diffusion period of duration $\tau_d$ consist of simultaneous application of pulses at $E_y$ ($f_\mathrm{RO}$) and $E_{\pm1}$ ($f_\mathrm{SP}$) for $100\,$\textmu s at 5 nW. We consider correlation forward and backward in time, by only selecting $C_2$ counts for repetitions where $C_1>C_{th}$ and vice-versa. We set $C_{th}=40$, giving selection probabilities in the 2-4\% range ($N=40000$).
     g) Data for a $P=2.5\,$\textmu W diffusion pulse at $f_\mathrm{RO}$. Solid lines are from a model assuming O-U diffusion of the spectral frequency (backward) and additionally frequency-dependent ionisation (forward). h) The extrapolated diffusion timescale $\tau_c$ scales as $P_d^{-1.05(4)}$. Blue diamond corresponds to data shown in (g). i) In the dark, no significant diffusion is observed.}
    \label{fig:optical}
\end{figure*}
For applications that use the NV centre as a photon source, such as quantum networks, a coherent optical transition with limited spectral diffusion due to charge noise is required \cite{hermans_entangling_2023,beukersRemoteEntanglementProtocolsStationary2024}. Here we investigate the same NV centre that was used for the CPMG measurements (Fig. \ref{fig:dd}), and show that our diamond growth yields good optical properties. Note that previous investigations of isotopically purified diamond observed increased spectral diffusion in comparison to natural abundance diamond, potentially caused by the presence of additional impurities \cite{bradley_robust_2022, degen_entanglement_2021, ishikawaOpticalSpinCoherence2012}.

First, we determine the optical level structure via a Photo-Luminescence-Excitation (PLE) scan (Fig. \ref{fig:optical}a). A 515~nm repump pulse is followed by a probe pulse at frequency $f_p$, while microwaves are applied to mix the $m_s=\pm1$ and $m_s=0$ states. The spectrum (Fig. \ref{fig:optical}b) matches the known NV-centre level system and confirms that the NV is in a low lateral strain regime ($V_E = 2.1\,\mathrm{GHz})$, comparable to values achieved in (100) CVD-grown and HPHT diamonds~\cite{mccullianQuantifyingSpectralDiffusion2022a, Blinder2024}. 

Secondly, we determine the linewidth of the $E_y$ optical transition. The lowest attainable (Fourier-limited) linewidth is set by the excited-state lifetime, but fluctuating charge traps arising from crystal defects can cause spectral diffusion, effectively broadening the optical transition. We use check-probe measurements to determine the linewidth \cite{vandestolpeCheckprobeSpectroscopyLifetimelimited2025} (Fig. \ref{fig:optical}c). First, a check pulse at frequency $f_c$, resonant with the $E_y$ transition, checks the charge state and frequency resonance of the emitter, which are heralded by the number of check counts $C$ exceeding a set threshold $C_{th}$. Subsequently, a detuned probe pulse at $f_p$ probes the response of the emitter, yielding $P$ probe counts. For $f_p=f_c$, this is similar to the common charge-resonance check used to prepare the charge state and frequency resonance of defect centres prior to experiments \cite{bernien_heralded_2013, brevoord_heralded_2023}. The measurement is repeated for different $f_p$ to determine the linewidth. CW microwaves are applied during laser pulses to mix the ground state spin levels, to avoid spin pumping.

Setting no threshold ($C_{th}=0$) yields the inhomogeneous linewidth $\gamma_i=222(3)$ MHz (FWHM, Fig. \ref{fig:optical}d). This linewidth arises from charge scrambling induced by the repump pulse prior to each measurement. Thresholding at $C_{th}=30$ approaches the homogeneous line with FWHM $\gamma_h= 16.9(4)$ MHz, close to the NV-centre lifetime-limited linewidth of $\sim13$~MHz and among the lowest values measured in literature (Fig. \ref{fig:optical}e) \cite{hermans_entangling_2023,orphal-kobin_optically_2023}. 
With charge-resonance checks, spin-selective excitation gives an average NV single-shot readout fidelity of 92.5\%~\cite{robledo_highfidelity_2011}. We note that this single-shot readout was also used to enable the long spin coherence measurements in section \ref{sec:spin}.

To characterize the spectral stability after frequency resonance is heralded, we measure the diffusion rate during optical excitation (Fig. \ref{fig:optical}f) \cite{vandestolpeCheckprobeSpectroscopyLifetimelimited2025, bownessLaserInducedSpectralDiffusion2025}. Here, two ``check" blocks ($C_1$ and $C_2$) are separated by a diffusion period with duration $\tau_{\mathrm{d}}$, during which a laser at $f_\mathrm{RO}$ is applied to induce diffusion. The check blocks consist of simultaneous application of lasers at $f_{\mathrm{RO}}$ and $f_{\mathrm{SP}}$ to continuously excite all ground-state spin states. A high number of $C_1$ counts heralds frequency resonance of the emitter. When spectral diffusion occurs during the diffusion period, this will cause the subsequent number of $C_2$ counts to be smaller.  The time scale at which significant diffusion occurs while the readout laser is applied determines how many coherent photons the defect centre can emit before the frequency resonance has to be re-prepared. Both correlations forward in time and backward in time can be considered, by post-selecting $C_2$ counts conditioned on $C_1>C_{th}$ ($C_2[C_1>C_{th}]$), and vice-versa ($C_1[C_2>C_{th}]$). This separates ionization, which occurs only in forward-time, and diffusion, which is a time-symmetric process \cite{vandestolpeCheckprobeSpectroscopyLifetimelimited2025}.

The data reveals a relatively time-symmetric decrease in photon counts, a signature of spectral diffusion dominating over ionisation (Fig. \ref{fig:optical}g). The photon counts tend to a non-zero value for long $\tau_d$, in contrast to recent experiments on the $V_2$ defect in SiC where a decay to 0 counts was observed \cite{vandestolpeCheckprobeSpectroscopyLifetimelimited2025}. This is caused by a relatively small ratio of $\gamma_i/\gamma_h$; The inhomogeneous linewidth limits the allowed frequency range of the spectral diffusion, so for smaller $\gamma_i$ there remains a probability for the transition frequency to be close enough to $f_\mathrm{RO}$, even for long $\tau_d$.

To account for this difference, we adapt previous models \cite{vandestolpeCheckprobeSpectroscopyLifetimelimited2025} by describing the spectral diffusion as an Ornstein-Uhlenbeck (O-U) process \cite{bownessLaserInducedSpectralDiffusion2025}, which describes a random walker confined to a harmonic potential. The model predicts the probability distribution $P(f, \tau_d)$ of the transition frequency $f$ over time. The dynamics are determined by two parameters: the diffusion parameter $D$, which is related to the timescale at which diffusion occurs, and the inhomogeneous linewidth $\gamma_i$, which determines the frequency range that the emitter can diffuse over. This model is chosen because it reproduces the Gaussian steady-state probability distribution for $\tau_d\rightarrow\infty$: the inhomogeneous spectral lineshape. In O-U diffusion, $P(f,\tau_d)$ is a Gaussian distribution with a time-dependent variance \begin{equation}
    V(\tau_d) = \frac{\gamma_i^2}{8\ln(2)}\left(1-\exp\left[\frac{-2D \tau_d}{\gamma_i^2/(8 \ln(2))}\right]\right).
\end{equation} To relate $D$ to a more easily interpretable physical quantity, we define $\tau_c$ as the diffusion duration for which the FWHM of $P(f,\tau_c)$ is equal to the homogeneous FWHM $\gamma_h$; this sets a rough timescale at which the diffusion significantly affects the optical linewidth. $\tau_c$ and $D$ are related via\begin{equation}
    \tau_c\approx\frac{\gamma_h^2}{16\ln(2)D},
\end{equation} assuming $\gamma_h^2/\gamma_i^2\ll 1$. 
Our model first calculates $P(f,\tau_d)$, which is then convolved with a Lorentzian of height $C_0$ and linewidth $\gamma_h$, corresponding to the homogeneous response, to calculate the expected number of counts $C(\tau_d)$. We fix $\gamma_h=22$~MHz, which is set by the power broadening during the $C_1$ and $C_2$ pulses (see \ref{sec:app_OUmodel} for more details). 

We perform the diffusion measurement for a range of diffusion-pulse powers $P_d$. We perform a joint fit of the model to the backward-time correlation data for all $P_d$, fitting a single $\gamma_i$, with separate $D$ and $C_0$ for each $P_d$ (Fig. \ref{fig:optical}g). This choice is motivated by the expectation that the inhomogeneous linewidth is determined by the number of fluctuating charge traps in the environment, which is expected to primarily depend on the photon energy, not the laser intensity. The model fits well to the data (reduced chi-squared $\chi_r^2=2.1$), indicating that the assumption of a power-independent $\gamma_i$ is reasonable (see \ref{sec:app_OUmodel} for data at all powers). 
The diffusion timescale scales as $\tau_c\propto P_d^{-1.05(4)}$ (Fig. \ref{fig:optical}h). At $P_d=15$~nW (the saturation power when a 1~\textmu s pulse is applied), the fit predicts $\tau_c>0.8$~ms. Recent measurements on T-centres \cite{bownessLaserInducedSpectralDiffusion2025} and erbium atoms \cite{fruhSpectralStabilityCavityenhanced2026} in silicon show a similar scaling of spectral diffusion with $P_d$. %
Without an applied laser, no spectral diffusion is detected within one second (Fig. \ref{fig:optical}i).
The fit inhomogeneous FWHM $\gamma_{i,\mathrm{red}}=117(2)$~MHz from our model corresponds to the inhomogeneous linewidth due to the diffusion laser (near 637~nm). This is smaller than $\gamma_i$ under 515~nm excitation (222(3)~MHz, Fig. \ref{fig:optical}d), which is likely caused by the lower energy 637~nm light exciting a smaller number of charge-trap species. Under 515~nm excitation we also observe diffusion timescales that are an order of magnitude faster at the same laser power, implying that $\tau_c$ also has a photon-energy dependence. 

To also model the forward-time data we consider the $\mathrm{NV}^-\rightarrow \mathrm{NV}^0$ ionisation rate, which is caused by resonant $f_\mathrm{RO}$ laser light. We use the same diffusion model and parameters as for the backward-time fits. First of all, we adjust the model counts for an overall slight decrease (4\%) in the forward-time counts for all $P_d$, caused by ionisation during the $C_1$ and $C_2$ pulses. For ionisation during the diffusion period we develop an additional model, which takes into account that ionisation only occurs when the emitter is resonant with the diffusion laser. Because there is no closed-form solution for O-U diffusion with frequency-dependent ionisation, our model instead performs an asymptotic approximation of $P(f,\tau_d)$, assuming a $\delta$-function ionisation rate when $f=f_\mathrm{RO}$ \cite{chowFokkerPlanckEquation2004} (see \ref{sec:supp_ion}).

The methods described here enable quantitative comparison between emitters in different diamonds. To put the obtained values in perspective, we perform the same spectral diffusion measurements on an NV centre that was previously used in remote entanglement experiments \cite{stolkMetropolitanscaleHeraldedEntanglement2024} (\ref{sec:qne}), and is hosted in a non-isotopically-purified (100)-grown diamond (Element Six). We find a smaller $\gamma_{i,\mathrm{red}}=42(9)$ MHz and values for $D$ that are $\sim30$ times lower. This comparison to the state-of-the-art for (100)-oriented growth shows that, while our (111)-oriented growth results in stable lines with limited spectral diffusion, further improvements in material purity might be possible. Reducing the diffusion rate would, for example, enable more repetitions of remote entanglement protocols before requiring a new charge-resonance check, which can increase the overall entanglement rate.   
We hypothesize that in our (111)-grown sample, the spectral diffusion under 637 nm excitation is caused by photoionisation of single vacancies (GR1, photoionization at 1.67 eV \cite{walkerOpticalAbsorptionLuminescence1979,kiflawiElectronIrradiationFormation2007}) and vacancy complexes that were incorporated in the CVD film during growth. The extra diffusion during 515 nm excitation is likely caused by the remaining $\mathrm{N_s}$ centres (photoionization at 2.2 eV \cite{walkerOpticalAbsorptionLuminescence1979, rosaPhotoionizationCrosssectionDominant1999, orphal-kobin_optically_2023}).

\section*{Discussion}
We have presented NV centres in isotopically engineered diamond with electron-spin coherence times up to 11.2(8)~s, in combination with stable optical transitions. These results are made possible by: (1) a high-purity, low-strain, (111)-oriented diamond growth process with a precisely controlled $^{13}\mathrm{C}$ concentration; (2) spin-echo sequences that characterize and mitigate 50-Hz related noise, and (3) quantitative methods to characterize the spectral diffusion. 

Our results open several new avenues. The quantitive analysis of spectral diffusion enables systematic comparisons of optical coherence between different emitters and material platforms, and the long electron-spin coherence can be used to precisely probe spin impurities. Together with future investigations into the nature of the impurities responsible for the spectral diffusion, for example by measuring frequency dependencies \cite{orphal-kobin_optically_2023}, or by directly probing dark electron spins \cite{marcksQuantumSpinProbe2024, pieplowQuantumElectrometerTimeresolved2025}, our methods can inform further improvements in high-purity material growth.

In addition, our results provide further insight into the electron-spin coherence and the degree of isotopic purification that still yields practical benefit. When coherence starts to be limited by factors other than the nuclear spin bath, further isotopic purification primarily reduces the number of $^{13}\mathrm{C}$ spins that can be used as qubits, for example as memory qubits \cite{Bradley2022NPJQI}. In addition, the initially observed limit to $T_2^\mathrm{Hahn}$ and $T_2^*$ values for low $^{13}\mathrm{C}$ concentrations, and the consistency with which these are found in the literature \cite{bar-gillSolidstateElectronicSpin2013, balasubramanianUltralongSpinCoherence2009, ishikawaOpticalSpinCoherence2012, herbschlebUltralongCoherenceTimes2019, andersonFivesecondCoherenceSingle2022, bergeronSiliconIntegratedTelecommunicationsPhotonSpin2020}, highlights the importance of characterizing and mitigating 50~Hz related noise. Further improvements in spin coherence will likely be based on isotopic purification in combination with reducing external magnetic noise (e.g. by magnetic shielding) and spin impurities in the diamond.

The developed high-purity (111)-oriented growth in isotopically engineered diamond might create new opportunities for quantum simulations of many-body physics, where coupled spin ensembles with preferential alignment are beneficial \cite{PhysRevX.15.021035}. Finally, the record coherence times demonstrated might lead to further improved spin-qubit control for quantum networks based on colour centre qubits \cite{bartling_universal_2024, pompili_realization_2021}.

\section*{Methods}
\label{sec:methods}
\paragraph*{\textbf{Diamond growth setup}}
All diamond epilayers were grown using a
MPCVD system (SDS6300, Seki Diamond Systems, Cornes Technologies Ltd.) equipped with a 2.45\,GHz magnetron and a turbomolecular pump.
Diamond substrates were mounted on a cylindrical molybdenum holder (12\,mm diameter with a central counterbore), ensuring a run-to-run variation of the diamond surface height relative to the holder surface within $\pm100~$\textmu m.
The surface temperature was controlled by the microwave input power ($<500$\,W) and monitored in situ through a viewport using an optical pyrometer DWF-24-40-C (Williamson), or DFP-2000 (Spectrodyne Inc.).

Process gases consisted of hydrogen ($>$99.999\%), methane with natural $^{13}$C abundance ($^{13}$C = 1.1\%, purity $>$99.9995\%), and isotopically enriched methane
($^{12}$C $>$99.99\%, purity $>$99.9\%; Cambridge Isotope Laboratories).
Gas purifiers PS7-PD05 for H$_2$ and PS4-MT3 for CH$_4$ (SAES Pure Gas Inc.) were used to minimize residual impurities, such that the nitrogen concentration in the mixed gas was expected to be a few \,ppb or lower, based on the manufacturer specifications and the low methane concentration (CH$_4$/H$_2 <0.1$\%).

\paragraph*{\textbf{Isotopically controlled samples}}

Undoped and isotopically engineered diamond epilayers (samples \emph{Fukuoka}, \emph{Kagawa}, and the calibration sample \emph{LD3}) were grown homoepitaxially on low-dislocation high-pressure high-temperature (HPHT) type-IIa diamond substrates.
The HPHT substrates (Technological Institute for Superhard and Novel Carbon Materials) were used to reduce dislocation density and background
photoluminescence from substrate NV centers.
The concentration of P1 centres (substitutional nitrogen) in the substrates, estimated by electron spin resonance, was approximately 1\,ppm.

Hydrogen flow rates were set to 300\,sccm for \emph{Fukuoka} and 2000\,sccm for \emph{Kagawa} and \emph{LD3}.
Methane flow rates were controlled using mass-flow controllers FCST1000MF series (Fujikin Inc.), yielding CH$_4$/H$_2$ ratios of approximately 0.03\% for
\emph{Fukuoka}, 0.08\% for \emph{Kagawa}, and 0.01--0.08\% for \emph{LD3}.
Growth was performed at a total pressure of 120\,Torr and a surface temperature of $915\pm25\,^\circ$C with microwave power below 500\,W.

\paragraph*{\textbf{Intentionally nitrogen-doped samples}}

Three intentionally nitrogen-doped diamond samples (S1--S3) were grown on Sumicrystal$\textsuperscript{\textregistered}$ HPHT type-Ib diamond substrates (Sumitomo Electric).
Growth conditions were fixed at CH$_4$/H$_2$ = 0.05\%, a total pressure of 100\,Torr, and a microwave power of 420--490\,W.

Samples S1 and S2 were grown under identical conditions, each containing two nitrogen-doped layers formed with gas-phase N$_2$/CH$_4$ ratios of $1.2\times10^5$ and $5\times10^5$\,ppm.
These conditions yielded nitrogen concentrations of 8 and 47\,ppm in S1, and 7 and 43\,ppm in S2.
Sample S3 was grown with a gas-phase N$_2$/CH$_4$ ratio of $1.0\times10^6$\,ppm, resulting in a nitrogen concentration of 72\,ppm.

\paragraph*{\textbf{Isotope quantification and calibration}}

Carbon isotope compositions are conventionally expressed as the $^{13}$C/$^{12}$C ratio relative to the Vienna Pee Dee Belemnite (VPDB) standard~\cite{VPDB}.
Throughout this work, we use the $^{13}$C concentration
$\chi = {}^{13}\mathrm{C}/({}^{12}\mathrm{C}+{}^{13}\mathrm{C})$;
$\chi$ is related to the isotope ratio
$R = {}^{13}\mathrm{C}/{}^{12}\mathrm{C}$ via
$\chi = R/(1+R)$.

To place $\chi$ on an absolute scale, a multi-layer calibration structure with different $^{13}$C concentrations (sample LD3) was analysed by SIMS.
To avoid possible bias from gas-memory effects during switching between natural-abundance and $^{12}$C-enriched methane, SIMS ratios were not referenced directly to VPDB.
Instead, instrumental offsets were removed using $A = R_{\rm VPDB}/R_{\rm Ref}$, where $R_{\rm VPDB}=0.011113\pm0.000022$ is the IUPAC-recommended value for VPDB~\cite{VPDB}, and $R_{\rm Ref}$ is the SIMS isotope ratio of an external natural-abundance type-IIa $\{100\}$ CVD diamond reference.
The $^{13}$C concentration of the calibration sample LD3 layers was then obtained as
\[
\chi = \frac{A \cdot R_{\rm LD3}}{1 + A \cdot R_{\rm LD3}} .
\]

The natural-abundance layers of sample LD3 ($\chi_{1}=1.0937(86)\%$) were compared with an external natural-abundance $\{100\}$ reference diamond using the isotope delta notation~\cite{BrandCoplenVoglRosnerProhaska+2014+425+467},
\[
\delta_{\rm LD3/Ref}
=
\left(\frac{R_{\rm LD3}}{R_{\rm Ref}} - 1\right)\times 1000
\approx -5~\text{\textperthousand}.
\]
This small negative offset may arise from gas-memory effects during methane switching, variations in the isotopic composition of the natural-abundance methane supply, SIMS-related systematics, or a combination thereof.
Even if the full magnitude of the observed offset were solely attributed to gas-memory-induced mixing and were to affect the $^{12}$C-enriched layers in an analogous way, it would induce an offset of equal magnitude and opposite sign in $\chi_0$, corresponding to a shift of at most $\Delta \chi_0 \approx +0.06$~ppm ($\chi_0 \approx 13.06$~ppm).
This is negligible compared with the $\pm3$~ppm measurement uncertainty.
We therefore find no evidence for a measurable gas-memory effect and quote $\chi_0 = 13(3)$~ppm as our best estimate on the VPDB scale.

For completeness, the isotope delta of the $\{100\}$ reference diamond relative to VPDB is
\[
\delta_{\rm Ref/VPDB}
=
\left(\frac{R_{\rm Ref}}{R_{\rm VPDB}} - 1\right)\times 1000
= +13.2~\text{\textperthousand}.
\]

\paragraph*{\textbf{Nanofabrication}}
Nanofabrication of SILs and striplines was performed on Fukuoka. First, room-temperature pulsed ESR was used to identify spatially isolated NV candidates. SILs were fabricated using a dual-beam Focussed Ion Beam (FIB) setup. High-power laser exposure of the surface during the optical characterization was used to leave recognizable markings under electron-beam microscopy, to align in the dual-beam FIB setup. Because the surface exhibited recessed pit-like features with a terraced structure, likely formed by crystallographically anisotropic etching and subsequent overgrowth during step-flow growth,
a novel method to pattern gold was used.
The striplines were patterned using direct laser lithography, aligning on the SILs, with negative resist and liftoff. The gold was evaporated under an angle on a rotating platform to ensure step-coverage. Direct laser lithography and liftoff of an ALD deposited $\sim100$ nm AlO$_x$ layer was performed to act as an anti-reflective coating, but this layer was damaged during lift-off.

\paragraph*{\textbf{NV Experimental Setups}}
Measurements of bulk NV centres on both Kagawa and Fukuoka (Fig. \ref{fig:spin}a-c, grey data in d) were performed on a first setup, while all other NV spin and optical measurements were performed on a second setup on an NV in a SIL in Fukuoka. Both setups consist of a Montana Cryostation S50 to cool the diamond to 4~K, with a confocal microscopy setup for optical collection (avalanche photon detector) and excitation (AOMs used for pulse shaping). Microwave pulse generation is performed with a ZI HDAWG arbitrary waveform generator and an R\&S SGS100A MW source. For bulk NVs square pulses were used, for the SIL NV hermite-shaped pulses were used. After MW amplification a MW switch is used to minimize noise that may flip or dephase the electron spin \cite{abobeih_onesecond_2018}. The primary difference between the setups is their location in the building, the position of equipment with respect to the cryostat, and the usage of inner-outer DC blocks in the second setup to mitigate ground loops, which may explain the difference in Hahn-echo signal between the setups (Fig. \ref{fig:spin}d), and the slightly longer $T_2^*$ in the second setup (Fig. \ref{fig:spin}g). All data was taken using the QMI Python package \cite{teraaQMIQuantumMeasurement2026}.

All spin measurements consisted of a (1) charge-resonance (CR) check \cite{bernien_heralded_2013}, (2) a laser pulse resonant to the $E_{\pm1}$ transition to initialize the ground-state spin, (3) a sequence of MW pulses, (4) a laser pulse resonant to the $E_y$ or $E_x$ transition to readout out the spin state in a single shot. Typical average single-shot readout fidelities \cite{robledo_highfidelity_2011} in bulk lie around 65\% (in both samples), for the SIL NV in Fukuoka this is up to 92.5\% (20 nW laser power, 8 $\mu$s pulse length). Spin-state readouts are corrected for the known SSRO infidelity. 

For the CPMG data the phase of the final $\pi/2$ pulse was alternated between 0 and $\pi$ and the difference between the two resulting estimated electron-spin-state populations was used to correct for drifts in photon collection efficiency. For presentation purposes, the CPMG data and fits (Fig. \ref{fig:dd}b and \ref{sec:dd_complete}) were rescaled by the fit parameter $A$, which ranges from 0.9 to 1.1. The Ramsey data of bulk NVs (Figs. \ref{fig:spin}a-c) was measured using slow $\pi/2$ pulses, so that an oscillation only occurs when the nitrogen spin is in the $m_I=0$ state. For the SIL NV (Fig. \ref{fig:spin}g), fast $\pi/2$ pulses were used, and the data was post-processed with a 100~kHz low-pass filter to eliminate the $\sim2$~MHz oscillations due to the $m_I=\pm1$ states.

The magnetic field for bulk measurements on Fukuoka (Kagawa) was $\sim1.5$ mT ($\sim2.5$ mT), and for measurements on the SIL NV in Fukuoka it was $\sim3.6$ mT. Magnetic fields were applied with an external permanent magnet and aligned along the surface-normal $\left[111\right]$ NV (`$z$') axis. The field strengths are large enough to place the NVs in the strong-field regime in which perpendicular hyperfine couplings of the nuclear spins may be ignored in the Ramsey experiments \cite{zhao_decoherence_2012}. The magnetic field required for this regime scales as $\chi$ because the typical nuclear spin couplings similarly scale as $\chi$, and is therefore much lower than for natural abundance. 

\normalsize
\section*{Acknowledgements}
We thank Yuhei Sekiguchi and Hideo Kosaka for performing preliminary low-temperature PLE measurements on diamond samples.
We thank Satoshi Koizumi and Philippe Bergonzo for fruitful discussions on diamond growth. We also thank Ryoichi Ishihara for selecting the CVD-related equipment, and Charles de Boer for assisting with its installation in the Kavli Nanolab cleanroom at Delft University of Technology. We thank Breno Perlingeiro Correa and Georgios Delaroudis for their contribution to the nanofabrication on the Fukuoka sample. We thank Peter Schaeffers for his assistance in preliminary room-temperature characterisation of the NV samples. We thank Pieter Botma for preparing the spectral diffusion measurements for the NV center from \cite{stolkMetropolitanscaleHeraldedEntanglement2024}. We thank Régis Méjard for maintenance support on one of the low-temperature measurement setups. We thank Mohamed Abobeih and Guido van de Stolpe for discussions.

We acknowledge funding from the Dutch Research Council (NWO) through the project “QuTech Phase II funding: "Quantum Technology for Computing and Communication” (Project No. 601.QT.001). This project has received funding from the European Union’s Horizon Europe research and innovation program under grant agreement No 101135699. We gratefully acknowledge support from the joint research program “Modular quantum computers” by Fujitsu Limited and Delft University of Technology, co-funded by the Netherlands Enterprise Agency under project number PPS2007. This work was supported by the Netherlands Organization for Scientific Research (NWO/OCW), as part of Quantum Limits (project number SUMMIT.1.1016). We also acknowledge support from the Japan Science and Technology Agency (JST), Moonshot R\&D Program (JPMJMS2062).

\section*{Contributions}
T.Y. conducted the diamond growth experiments, while B.v.O. and K.-N.S. performed the optical and spin measurements.
B.d.Z. assisted in preliminary room-temperature characterisation of the NV samples. 
S.S. measured the nitrogen concentration in the type IIa HPHT substrate, and S.O. and T.O. carried out the electron beam irradiation experiments.
R.V. and H.A.-T. fabricated the SIL device. %
T.Y. and B.v.O. wrote the paper, with input from T.H.T. and K.-N.S.. 
T.H.T. supervised the overall research.

\section*{Data Availability}
The data and analysis code underlying this study are available on the open 4TU data server \cite{dataserver}.

\bibliography{refs}	
\clearpage
\appendix
\newcounter{chapter}
\renewcommand{\thechapter}{\Alph{chapter}}
\setcounter{chapter}{19}
\renewcommand{\thesection}{Supplementary Note \arabic{section}}
\setcounter{section}{0}

\renewcommand{\theequation}{S\arabic{section}.\arabic{equation}}
\setcounter{equation}{0}

\renewcommand{\thefigure}{S\arabic{figure}}
\setcounter{figure}{0}

\renewcommand{\thetable}{S\arabic{table}}
\setcounter{table}{0}
\counterwithin{figure}{chapter}
\begin{widetext}
\begin{center}
\textbf{\large Supplementary Information for \\``Ten-second electron-spin coherence in isotopically engineered diamond''}
\end{center} 
\normalsize
\renewcommand{\appendixname}{}

\section{Nitrogen sources: leak versus outgassing}
\label{sec:app_nitrogen_outgas}
\begin{figure}[h]
  \centering
  \includegraphics[width=8cm]{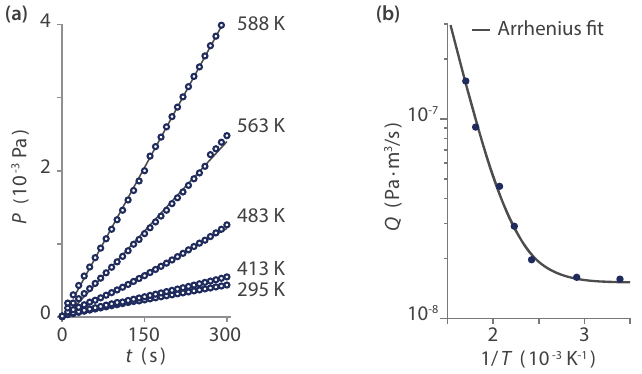}
  \caption{
\emph{Leak-rate characterization.}
(a) Pressure rise after pump-off measured at several temperatures as a function of time; linear fits yield the pressure rise rate $dP/dt$.
(b) Arrhenius plot of the effective gas throughput $\log Q = \log[V(dP/dt)]$ as a function of inverse temperature ($1/T$).
}
  \label{fig:leak}
\end{figure}

To quantify the nitrogen concentration in the reactor gas, we first quantify the gas influx, given by $Q=V(dP/dt)$. We measured the pressure rise rate $dP/dt$ in the reactor chamber after pump-off at sample-stage temperatures $T = 295$–$588$~K (Fig.~\ref{fig:leak}a). Here $V=11.3$ L is the effective reactor volume. $Q$ has a temperature-dependent and -independent component, corresponding to outgassing from chamber materials and vacuum leaks, respectively. We extract these by fitting an Arrhenius expression $Q =Q_{\rm leak} + Q_0 \exp[-E_a/(k_{\rm B}T)]$ (Fig.~\ref{fig:leak}b). Since nitrogen outgassing is expected to be negligible under our growth conditions~\cite{Dylla,Calder}, air leaking into the chamber is the primary source of nitrogen.
From the fit, we extract a temperature-independent leak component of $Q_{\rm leak} = (1.5\pm0.6)\times10^{-8}$~Pa\,m$^3$\,s$^{-1}$.

Under the growth pressure ($P_{\rm in} = 120$~Torr) and atmospheric pressure $P_{\rm atm}$, the effective leak throughput becomes
\begin{align}
\label{eq:leakrate}
Q_{\rm leak}
  \frac{P_{\rm atm} - P_{\rm in}}{P_{\rm atm}}
= 1.3(0.5)\times10^{-8}\ {\rm Pa\,m^3\,s^{-1}} .
\end{align}
Since air contains approximately 78\% N$_2$ by volume, this corresponds to a nitrogen inflow rate of \begin{align}\label{eq:nitrogen_flow_app}
    \dot n_{\rm N_2} = 4.0(1.6)\times10^{-12}~\mathrm{mol}\,\mathrm{s}^{-1}.
\end{align}
We assume that this leak-derived nitrogen supply is continuously present during growth. 

Possible additional nitrogen sources include chamber outgassing and impurities in the process gases. However, both are expected to be small compared to the leak-derived contribution. Previous vacuum studies show that, after air exposure and pump-down, the outgassed flux from stainless-steel and aluminium chambers is dominated by water ($\gtrsim85\%$), with the remaining fraction consisting mainly of H$_2$, CO, CO$_2$, and CH$_4$, and no significant contribution from N$_2$~\cite{Dylla}. Upon heating, the outgassing becomes hydrogen-dominated ($>99\%$), implying that all other species, including N$_2$, contribute at most at the sub-percent level~\cite{Calder}. Accordingly, molecular nitrogen is generally
treated as a leak-related species rather than an outgassing product in ultra-high-vacuum systems.

During growth, the diamond surface is heated to $\sim915\,^\circ$C by the plasma, while the temperature measured inside the sample stage increases to $\sim310\,^\circ$C due to plasma heating.
Because leak measurements cannot be performed with plasma present, the latter temperature was set using a radio-frequency heater. The Arrhenius analysis (Fig.~\ref{fig:leak}b) shows that the thermally activated component increases by about an order of magnitude near this temperature ($1/T \simeq 1.7\times10^{-3}$~K$^{-1}$), consistent with outgassing.
Even if this component contained N$_2$ at the sub-percent level, the corresponding nitrogen outgassing would remain at least an order of magnitude smaller than the continuously present leak-derived N$_2$ inflow, and it would decrease further under long-duration steady-state growth.
We therefore conclude that outgassing is not a significant nitrogen source under the present growth conditions.

Nitrogen impurities in the purified process gases are also not expected to dominate. Based on the purifier specifications and the low CH$_4$/H$_2$ concentrations, the residual nitrogen concentration in the mixed process gas is expected to be at the level of a few ppb or lower. Even assuming a conservative upper bound of $\sim$1~ppm residual nitrogen in the gas supply, the corresponding solid-state contribution is $<0.1$~ppb given our incorporation efficiency. This contribution therefore remains well below the leak-derived nitrogen influx.

Finally, Shimaoka \emph{et al.} investigated an intentionally oxygen-added MPCVD (H, C, N, O) system and reported a moderate enhancement of nitrogen incorporation (by less than a factor of two, with an optimum oxygen level of $\mathrm{O/H}\sim0.5\%$)~\cite{Shimaoka}.
In our nominally undoped growth, however, the leak-induced oxygen level is only 0.7--4.5~ppb relative to the hydrogen flow; thus, even if present, any oxygen-related effect on nitrogen incorporation should be negligible.

Based on these considerations, we treat the leak-derived N$_2$ inflow as the dominant nitrogen source and as a lower bound for nitrogen incorporation.

To compare the nitrogen flow-rate with MFC-controlled methane flow rates calibrated at standard conditions (0$^\circ$C and 1~atm, in sccm), the nitrogen inflow at $298$~K can be converted to standard conditions using the corresponding conversion factor $1.345\times10^{6}$~sccm/(mol/s).
This corresponds to a nitrogen leak flow of $5.4\times10^{-6}$~sccm (95\% confidence interval: $3.3$--$7.5\times10^{-6}$~sccm), expressed at standard conditions.

\section{Simulation of $T_2^*$ in a $^{13}$C bath}
\label{sec:app_t2star_sim}
We describe the simulations that were performed to calculate the $^{13}$C limited $T_2^*$, or free-induction decay (FID) time distributions for the NV centre electron spin at various $^{13}$C concentrations $\chi$. 

To calculate the effect of $^{13}$C-spins on $T_2^*$, we follow the approach of Zhao \textit{et al.} \cite{zhao_decoherence_2012}. Since the nuclear-nuclear coupling in the carbon-spin bath is much weaker than the electron-nuclear coupling, the carbon-spin bath can be considered to be non-interacting on the timescale of $T_2^*$. Thus, the bath is considered quasi-static and the contributions of individual spins can be considered separately. Additionally, we only consider the $z$-component of the hyperfine interaction between electron and nuclear spins $A_z$, as a sufficiently large magnetic field was applied along the $z$ direction in all experiments. Under these assumptions, the coherence $L_\mathrm{FID}(t)$ in an FID experiment can be shown to decay as \begin{align}
    L_{\mathrm{FID}}(t) \approx e^{-\Gamma_z^2 t^2/2},
\end{align}
 where $\Gamma_z^2=\sum_j(A_{j,z})^2/4$. $T_2^*$ follows directly from $T_2^* = \sqrt{2}/\Gamma_z$. We calculate the hyperfine interaction between the electron and spin $j$ as \begin{align}
    A_{j,z} = \frac{\mu_0\hbar\gamma_c\gamma_e}{4 \pi r_j^3}\left(3 \cos^2(\theta_j)-1\right),
\end{align}
with $\gamma_c$ and $\gamma_e$ the carbon-spin and electron-spin gyromagnetic ratios, respectively, $r_j$ the distance from the electron spin to carbon spin $j$, and $\theta_j$ the polar angle to spin $j$. Because of the low nuclear-spin concentrations studied in this work the contact hyperfine contribution may be neglected because it is very unlikely for a nuclear spin to be close enough for it to play a role.

To calculate the $T_2^*$ distribution, we generate $10^5$ spin baths for each $\chi$ under investigation. We do not constrain sampled nuclear-spin positions to the diamond lattice, as this will have a negligible influence at low $\chi$. To generate a specific nuclear-spin bath, we randomly sample $N$ positions ($r_j, \theta_j$) from a sphere with radius $r_{max}=45$~nm, which is chosen large enough for the distribution to converge at our investigated $\chi$. The amount of positions sampled is \begin{align}
    N = \frac{4\pi}{3} r_\mathrm{max}^3n_d\chi,
\end{align}
with $n_d=1.76\times10^{29}\,\mathrm{m}^{-3}$ the density of atoms in diamond. For each position $j$ we calculate $A_{j,z}$, which are used to calculate $T_2^*$ for a single bath configuration.

We note that the probability density function of $T_2^*$ approximately follows a half-normal distribution \begin{align}
    P(T_2^*)_\chi = \frac{\sqrt{2}}{ \sqrt{\pi} T_0/\chi}\exp\left[-\frac{(T_2^*)^2}{2(T_{0}/\chi)^2}\right],
\end{align}
with $T_0 \approx 0.0318\, $\textmu s.%

In principle, $r_{max}$ can be adjusted with $\chi$, as a denser bath will require a smaller volume to be simulated for the simulation to converge, but this optimisation was not performed. In addition, spins that are sufficiently strongly coupled to cause identifiable beating in a Ramsey signal could be eliminated from the $T_2^*$ calculation, which would cause a slight decrease in the probability for low $T_2^*$ \cite{marcksGuidingDiamondSpin2024}.

\section{Estimation of electron-spin concentration from $T_2^*$ data}
\label{sec:app_t2star_espin}

\begin{figure}
    \centering
    \includegraphics{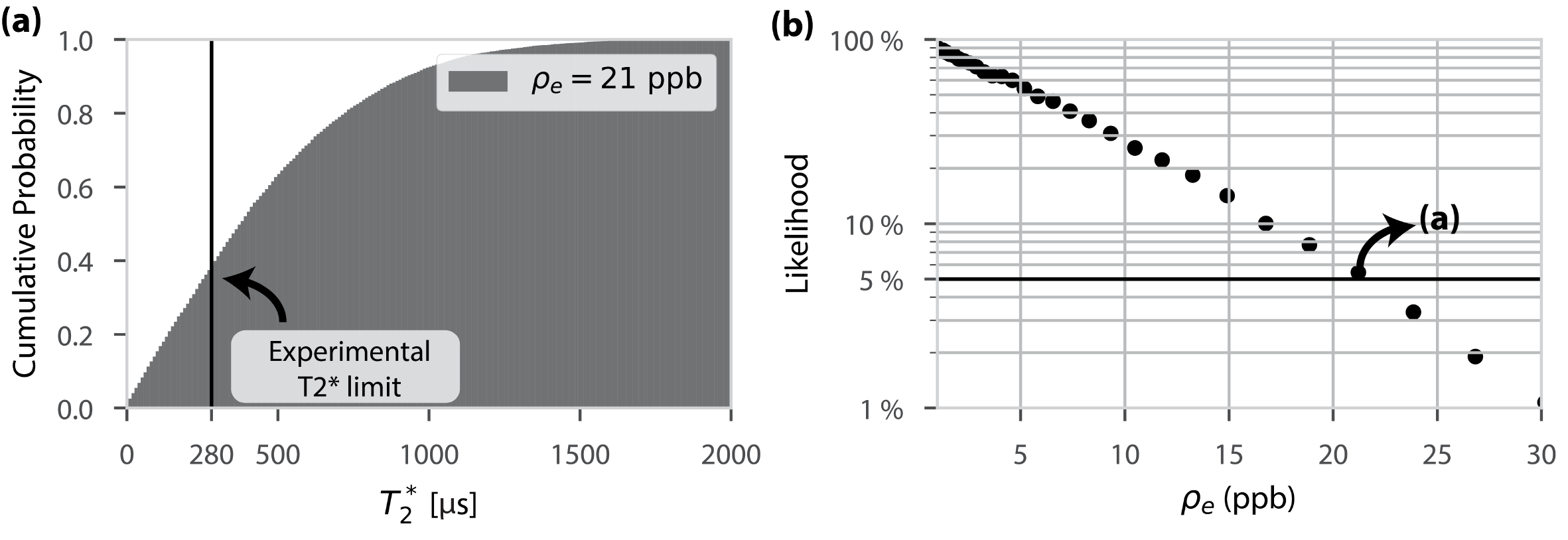}
    \caption{\emph{Estimating $\rho_e$ from $T_2^*$ measurements.} a) Cumulative probability distribution of $T_2^*$ for an NV centre in a bath of electron spins with a concentration of 21 ppb. The experimental $T_2^*$ limit is indicated (Fig. \ref{fig:spin}c). b) Likelihood of a given $\rho_e$, given that we measured 6 NV centres with a $T_2^*$ above 280~\textmu s. The likelihood of $\rho_e$ to be above 21ppb is less than $5\%$. }
    \label{fig:p1_t2star}
\end{figure}
In addition to using our knowledge of the growth process, and using SIMS or ESR methods, our measurements of NV-centre $T_2^*$ times can be used to estimate an upper bound on the concentration of electron spins $\rho_e$. This also puts an upper bound on the nitrogen concentration, as $P_1$ centres ($\sim75\%$ of $N_s$ \cite{Luo_2022,Teraji}) contribute to the electron-spin bath. %

The strongest upper bound on $\rho_e$ is given by our measurement of the $T_2^*$ of 6 NV centres in sample Fukuoka at $\chi=0.0013\%$ (Fig. \ref{fig:spin}c). It is unlikely that these measurements were limited by $\rho_e$, given the narrow distribution of $T_2^*$. However, let us assume that another noise source limits $T_2^*$ to approximately 280~\textmu s, and that the $^{13}\mathrm{C}$-bath may be ignored (which is a reasonable assumption as \C simulations predict a much longer $T_2^*$). For a given $\rho_e$, we can calculate the probability distribution of $T_2^*$ ($P(T_2^*|\rho_e)$), in a similar fashion as for $^{13}\mathrm{C}$ spins (Fig. \ref{fig:p1_t2star}a) \cite{zhao_decoherence_2012}. We are interested in the probability that among $N$ NV centres, each in an electron-spin bath with concentration $\rho_e$, no measured $T_2^*$ value lies below our observed lower limit $T_{2,L}^*$. Given the observation of $N$ NV centres with a lowest coherence time of $T_{2,L}^*$, we can calculate the likelihood for a given $\rho_e$ as \begin{align}
    L(\rho_e) = P(T_2^*>T_{2,L}^*|\rho_e)^N.
\end{align}
Our measurements show no $T_2^*$ value below 280~\textmu s for six measured NV centres. Given our simulations, this implies a likelihood for $\rho_e>21$ppb is less than 5\%. Our $T_2^*$ measurements thus give a rough upper bound of 21 ppb for the electron-spin concentration. The bound could have been made tighter by measuring more NV centres. 

The $\rho_e$ extracted from the $T_2^*$ measurements lies very closely to the $[N]<26$ppb upper bound extrapolated from the SIMS data, and so this method does not constrain our nitrogen concentration further. However, it does put a bound on the concentration of other electron spins arising from other defects in the diamond.

\section{Description of AC field sensing and 50Hz model}
\label{sec:app_50hz}
\begin{figure*}
    \centering
    
    \includegraphics{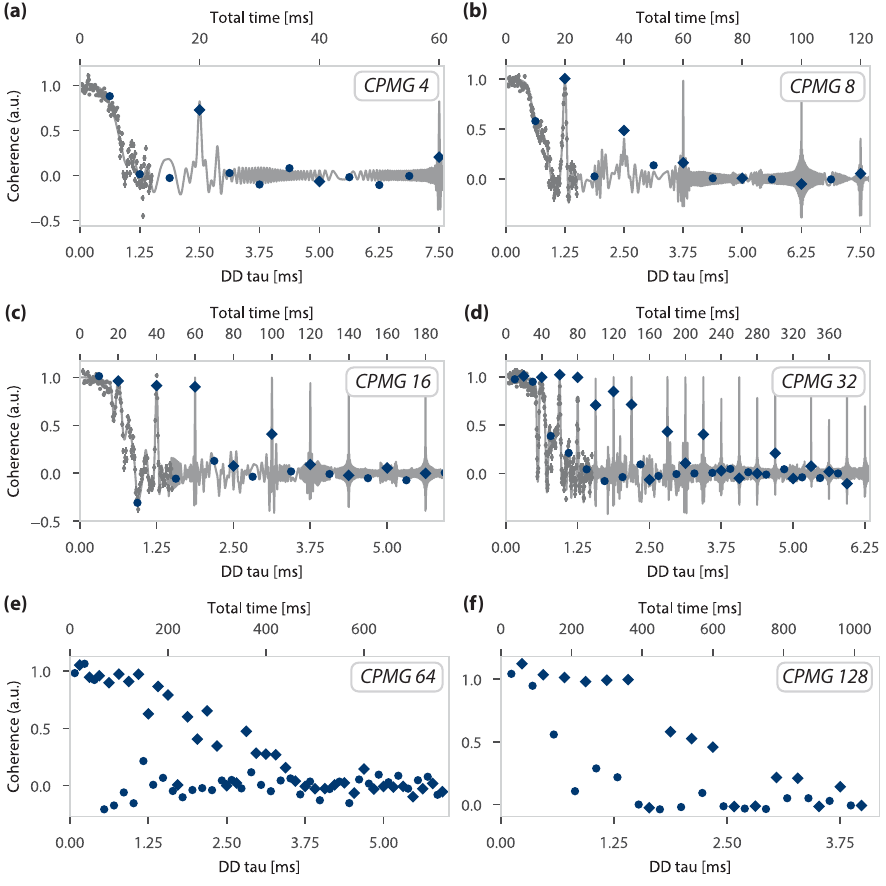}
    \caption{\emph{50~Hz noise in dynamical decoupling.} DD spectroscopy for 4(a), 8(b), 16(c), 32(d), 64(e), and 128(f) pulses. Up to $N=32$, we also plot the signal as predicted by the 50~Hz interference model (grey line). The experimental data (blue) is the same as in Fig. \ref{fig:dd}, however, for the "50~Hz mitigated" dataset, we show both data at multiples of 20~ms (triangles), and data offset by an additional 10~ms (circles). The data taken at a 10~ms offset show a very fast decay due to the 50~Hz interference. Not all revivals predicted by the model are present in the data. This is either caused by the fact that the mains frequency is not exactly 50~Hz during the measurement, or by other noise causing dephasing. The data at $\tau=2.5$~ms in the 4 and 8 pulse data is used to fix $B_{ac}$ for $f_{ac}=100$~Hz.}
    \label{fig:dd_hr_long}
\end{figure*}
This note summarizes well-known results based on the filter-function description of sensing sequences to obtain a model of the 50 Hz interference acting on the NV electron spin \cite{taylorHighsensitivityDiamondMagnetometer2008, ishikawaInfluenceDynamicalDecoupling2018}. We also discuss additional CPMG data showing the influence of the interference, and the conditions for the revivals observed in the CPMG data.

We assume the same magnetic field description as in the main text\begin{align}
    B_{ac}(t) = \sum_{i=1}^{n}B_i\cos (\omega_{i}(t-t_0) + \phi_i),
\end{align}
where the summation is over harmonics of 50 Hz (Table \ref{tab:50model}). Each field has its own magnitude $B_i$ and phase $\phi_i$, and there is a global time offset $t_0$. To describe the CPMG response of the electron spin, the phase that the spin picks up can be expressed as \cite{ishikawaInfluenceDynamicalDecoupling2018} \begin{align}
    \Phi_e = \gamma_{NV} \sum_{i=1}^{n} B_{i} \mathrm{Re}[ e^{-\mathbf{i} (\phi_{i}-\omega_it_0)} F(\omega_{i})],
\end{align}
where $\mathbf{i}$ is the imaginary unit, $\gamma_{NV}$ the NV gyromagnetic ratio (28 GHz/T), and $F(\omega_i)$ is the filter function of the CPMG sequence \cite{taylorHighsensitivityDiamondMagnetometer2008}: \begin{align}
     F(\omega_{i}) = T e^{-\mathbf{i} \omega_{i} N \tau}(1-\sec (\omega_{i} \tau))\frac{\sin(\omega_{i} N\tau)}{\omega_{i} N\tau},
\end{align}
with $T=2N\tau$ the total CPMG sequence time. Note that, while the filter function approaches a $\delta$-peak at $\omega=2\pi/(4\tau)$ for large $N$, we explicitly take into account the full form of $F(\omega)$ in this work.

The phase accumulated due to the sum of frequency components is then
\begin{equation}
    \Phi_{DD} = \sum_{i=1}^n\frac{2  \gamma_{NV} B_{i}}{\omega_{i}} \cos (\omega_{i}(N\tau - t_0) + \phi_{i}) (1 - \sec (\omega_{i} \tau) )\sin(\omega_{i}N \tau).
\end{equation}
The phase response for a Ramsey sequence of length $T$ and a Hahn-echo sequence with spin-echo time $2\tau$ can similarly be found by using \begin{equation}
    \Phi_{\mathrm{Ramsey}} = \gamma_{NV}\int_0^T B_{ac}(t)\mathrm{d}t = \sum_{i=1}^n \frac{2\gamma_{NV}B_i}{\omega_i}\sin(\omega_i T/2) \cos(\omega_i(T/2 - t_0)+\phi_i)
    \end{equation}
    \begin{equation}
    \Phi_\mathrm{Echo} = \gamma_{\mathrm{NV}}\left[\int_0^\tau B_{ac}(t) \mathrm{d}t  - \int_\tau^{2\tau} B_{ac}(t) \mathrm{d}t \right]
    = \sum_{i=1}^n\frac{4\gamma_{NV}B_i}{\omega_i}\sin^2(\omega_i \tau /2)\sin(\omega_i(\tau - t_0))
\end{equation}
The expected outcome of a single experiment repetition is given by the response for a single value of $t_0$ \begin{align}
    \left<X\right>_{t_0} = \cos\Phi_{DD}.
\end{align}

When the measurements are not synchronised to the mains frequency, the time $t_0$ will effectively be randomized for each repetition. The measured expectation value after many repetitions can be found be averaging over $t_0$ in the $[0, P]$ range
\begin{align}
    \left<X\right> = \frac{1}{P}\int_0^{P} \left<X\right>_{t_0} \mathrm{d}t_0,
\end{align}
with $P$ being the 20 ms period of the mains grid here. In our model, this was performed numerically by calculating $\left<X\right>_{t_0}$ for 400 evenly spaced values of $t_0$ and taking the average.

To create a model of the mains interference, we choose a discrete set of frequencies corresponding to $50$~Hz and harmonics up to $450$~Hz, and fit the amplitudes $B_i$ and phases $\phi_i$, fixing the phase of the $50$~Hz component to 0. We jointly fit the model to the higher-resolution $N=4,8,16,$ and $32$ data (from Fig. \ref{fig:dd}a, for parameters see Table \ref{tab:50model}). In figure \ref{fig:dd_hr_long} we plot the 50~Hz model and the high-resolution CPMG data it was fit to again, also including additional data from the long-$\tau$ dataset (from Fig. \ref{fig:dd}b) at decoupling times where no signal revival occurs. Diamonds correspond to predicted revivals and circles to non-revivals. This clearly shows that the signal decays rapidly when the revivals are not taken into account. The 100~Hz field amplitude and phase used in the model were fixed by looking at the $\tau=2.5$~ms data; at this specific $\tau$ only the 100~Hz component does not cancel out, as is explained next.

From the filter function description, the signal revivals observed in the CPMG data (Fig. \ref{fig:dd_hr_long}) can also be predicted. $F(\omega_{ac})=0$ occurs when both
\begin{align}\label{multiple}
    T &= k\ T_{ac} \ \forall\ k \in \mathbf{N}\\
    \land\ \tau &\neq \left(\frac{1}{4} + \frac{n}{2}\right) T_{ac}\ \forall\  n \in \mathbf{N},
\end{align}
with $T_{ac}=2\pi/\omega_{ac}$ the signal period. The first condition is straightforward: When satisfied by $f_{ac}=50$~Hz, it is also satisfied by any higher harmonic of 50~Hz. The second condition is more complex, and an exact analysis of when revivals occur in any situation is beyond the scope of the present discussion. It is relevant to mention that when $N$ is a power of 2, which applies to all data in this work, odd harmonics of 50~Hz (150~Hz, 250~Hz, ...) will also exhibit signal revivals when a 50 Hz signal would. On the other hand, even harmonics (100~Hz, 200~Hz, ...) remove some revivals. Our model also confirms this observation: revivals are predicted at $T=k\ 20$~ms, except at $\tau=5$~ms, when the second condition is not satisfied. At $\tau=2.5$~ms, the revivals are diminished because of the presence of 100~Hz noise, again because condition two is not satisfied for 100~Hz. The presence of other even harmonics is very weak, as can be seen by the strong signal revivals at $\tau=1.25$~ms.

\section{Additional Hahn-echo Data and $B_{ac}$ fluctuations.}
\label{sec:app_echoData}
\begin{figure}
    \centering
    \includegraphics{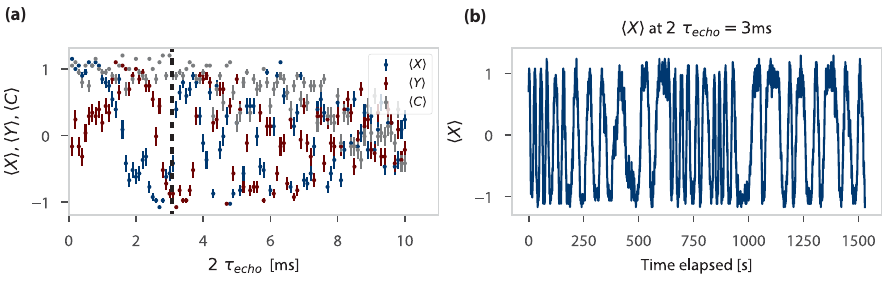}
    \caption{\emph{Additional synchronized $T_2$ data.} a) Representative example of a mains-synchronized Hahn-echo measurement (figs. \ref{fig:spin}d and e). While $\left< X \right>$ and $\left< Y \right>$ fluctuate a lot, a coherent signal $\left<C\right>$ can still be measured along the $\Phi_e$-axis. The average of $\left<C\right>$ over 12 such measurements results in the red trace in Fig. \ref{fig:spin}d. (b) A time trace of $\left< X \right>$ for a Hahn-echo sequence with $\tau_{echo}=1.5\, \mathrm{ms}$ (dashed line in (a)). On the second-timescale the outcome already fluctuates rapidly over time, underpinning that the mains-synchronized Hahn-echo measurement (a) must be performed within a few seconds for $\Phi_e$ to be corrected appropriately.}
    \label{fig:T2_sup}
\end{figure}
We provide extra data on the feedforward Hahn-echo measurement in figure \ref{fig:T2_sup}. To determine whether the amplitudes of the mains interference signals fluctuate over time, we repeatedly perform a synchronized Hahn-echo measurement with $\tau_{\mathrm{echo}}=1.5\, \mathrm{ms}$. The $\left<X\right>$ signal fluctuates on a second time-scale (Fig. \ref{fig:T2_sup}b), indicating that the amplitudes $B_i$ of the ac magnetic field fluctuate over time, which limits the efficacy of the feedforward scheme. 

The magnitude of the $B_{ac}$ fluctuations can be estimated from the time-trace data, by assuming that it can be explained a magnetic field with $B_{ac}'(t)=a(t)B_{ac}(t)$, where $B_{ac}(t)$ is the field as predicted by the model fit to the CPMG data (Table \ref{tab:50model}), and $a(t)$ represents the fluctuation of the ac amplitudes over time, with $E[a]=1$. We assume that the amplitudes fluctuate together. At $\tau_{echo}=1.5$~ms, the model prediction of $\left< X \right>$ as a function of $a(t)$ is well explained by $\left< X \right> =\cos (2\pi \times 1.18 a(t))$. Thus, a full swing from $\left<X\right>=1$ at $t_1$ to $\left<X\right>=-1$ at $t_2$ corresponds to $a(t_1)=0.85$ and $a(t_2)=1.27$. We use this fluctuation range to calculate the Ramsey signal envelope (Fig. \ref{fig:dd}g), by ensemble averaging the predicted envelope over this $a(t)$ range. We note that a small error is introduced because such fluctuations must also have been present when taking the CPMG data to which the $B_{ac}(t)$ model was fit, which causes the analysis to not be fully self-consistent. Nevertheless the qualitative behaviour should be properly captured by the above approach.

\section{Complete CPMG data}
\label{sec:dd_complete}
Some CPMG data was not presented in the main text to improve the legibility of the figure (Fig. \ref{fig:dd}b). We present all the data used to make the coherence scaling figure (Fig. \ref{fig:dd}c) here (Fig. \ref{fig:dd_complete}a), and include the fit stretch exponent $n$ (Fig. \ref{fig:dd_complete}b). A bound of 5 was set for $n$. For a bath of P1 centres, the expectation is $n=3$ \cite{desousaElectronSpinSpectrometer2009, delange_universal_2010}.
\begin{figure}
    \centering
    \includegraphics{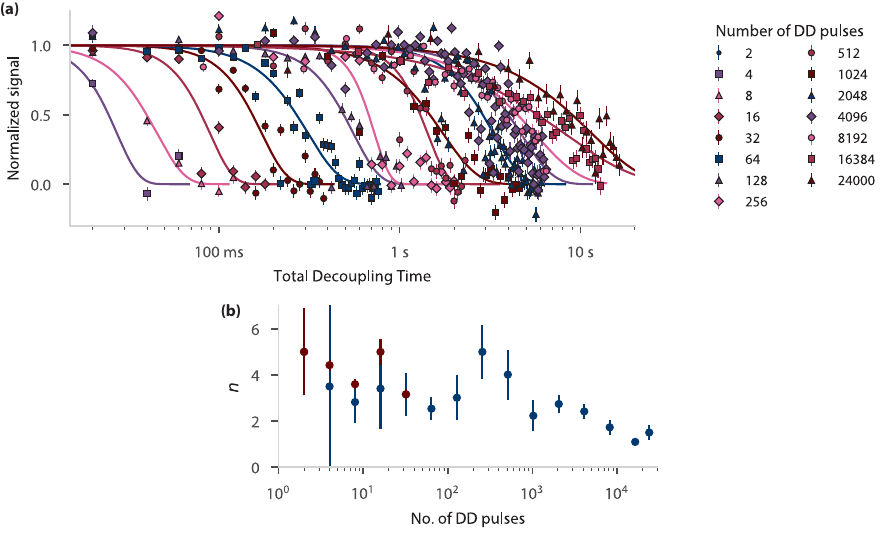}
    \caption{\emph{Full CPMG dataset.} a) CPMG dataset at predicted signal revivals. b) Stretch exponent $n$ (Eq. \ref{S_decay}) for the ``50Hz limited'' data (red) and ``50Hz insensitive'' data (blue).}
    \label{fig:dd_complete}
\end{figure}

\section{Comparison to CPMG coherence in other work}
\label{sec:app_CPMGcomp}
\begin{figure}
    \centering
    \includegraphics{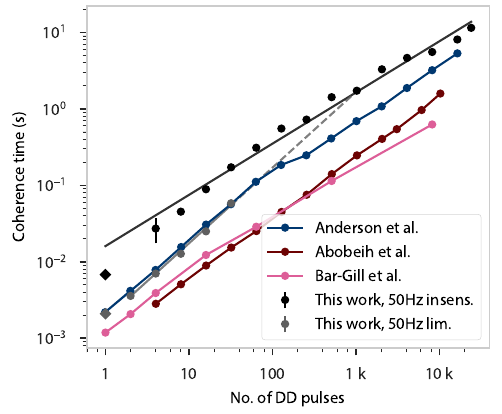}
    \caption{\emph{Comparison of $T_2^{DD}$ to other colour-centre work.} Anderson \textit{et al.} \cite{andersonFivesecondCoherenceSingle2022} measured a $V_2$ centre in isotopically purified SiC, Bar-Gill \textit{et al.} \cite{bar-gillSolidstateElectronicSpin2013} measured an NV centre ensemble in isotopically purified diamond, and Abobeih \textit{et al.} \cite{abobeih_onesecond_2018} measured a single NV centre in diamond with a natural abundance of ${}^{13}\mathrm{C}$.}
    \label{fig:t2_comp}
\end{figure}
We compare our CPMG coherence data to dynamical decoupling data from other studies in literature on defect centres in solid state (Fig. \ref{fig:t2_comp}). We plot data from Anderson \textit{et al.} (divacancy in Silicon-Carbide) \cite{andersonFivesecondCoherenceSingle2022}, Abobeih \textit{et al.} (NV centre in diamond, 4K) \cite{abobeih_onesecond_2018}, and Bar-Gill \textit{et al.} (NV centre, ensemble, 77K, isotopically purified) \cite{bar-gillSolidstateElectronicSpin2013}. The data from \cite{andersonFivesecondCoherenceSingle2022} roughly shows two scaling regimes, as also remarked by the authors; for low $N$, $\eta=0.92(1)$, but for higher $N$ there is a kink in the data to $\eta=0.75(1)$. Our 50~Hz-limited data closely follows the $\eta=0.92$ regime, hinting at a similar (60~Hz) limited regime in \cite{andersonFivesecondCoherenceSingle2022}. Data from \cite{bar-gillSolidstateElectronicSpin2013} possibly shows a similar kink, and for high $N$, follows a similar $\eta$ as our data.

We note the possibility of a similar kink being present in our ``50 Hz insensitive'' data from $N=4$ to $N\approx 100$. However, the first few coherence-time data points rely on fits to coherence values of only very few values for $\tau$ (Fig. \ref{fig:dd_hr_long}). This is because for small $N$ there are only few conditions for which $T=n\times 20$ ms before the electron spin decoheres (two for $N=4$, three for $N=8$), and these measurement points are further affected by 100 Hz noise at $\tau=2.5$ ms. Therefore it is not surprising that for low $N$ the extracted coherence time is shorter.

\section{Ornstein-Uhlenbeck model for spectral diffusion}
\label{sec:app_OUmodel}

In this note, we present the full spectral diffusion dataset (Fig. \ref{fig:optic_sup_diff}) and we derive the model used to fit the expected number of photon counts in the spectral diffusion experiments (Fig. \ref{fig:optical}f) to extract information about the nature and rate of spectral diffusion of the NV centre's optical transition. To model the photon counts, we first calculate the probability density function of the transition frequency $P(f,\tau_d)$. This is subsequently convolved with the homogeneous optical line to calculate the expected number of photons in the experiment.

$P(f,t)$ describes the probability for the optical frequency to be at frequency $f$ after a diffusion time $\tau_d$. We assume that before the diffusion period, the optical frequency is precisely at $f=0$: \begin{align}
    P(f, 0) = \delta(f),
\end{align} which can be approached experimentally by setting the selection threshold $C_{th}$ sufficiently high \cite{vandestolpeCheckprobeSpectroscopyLifetimelimited2025}.
\begin{figure*}
    \centering
    \includegraphics{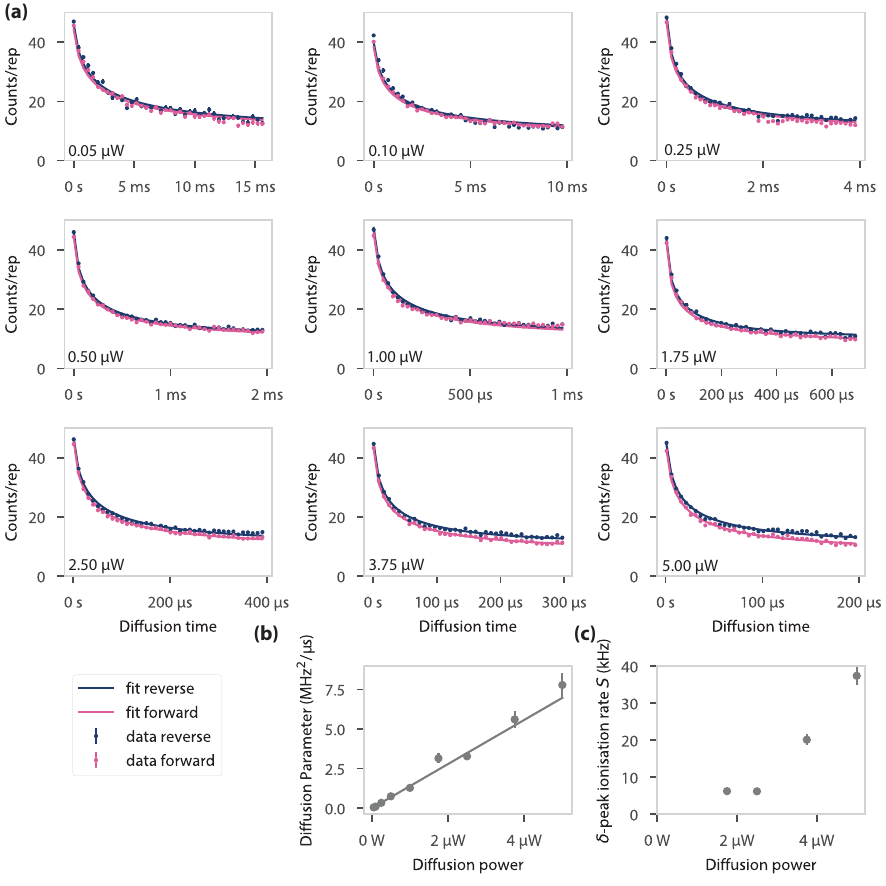}
    \caption{\emph{Diffusion data.} (a) Complete diffusion data set for all used diffusion laser powers, including fits by the O-U model. For the forward data above 1~\textmu W the diffusion model with ionisation was used to fit an ionisation rate. For all powers, the forward data is $\sim4\%$ lower for $1$ \textmu s of diffusion time (the first data point), which we attribute to ionisation of the emitter during the $C_1$ and $C_2$ block. We attribute the jump in the 0.25~\textmu W data around 2~ms diffusion time to a drifting microscope objective, as an overall decrease in the non-thresholded data is also observed there. (b) Diffusion parameter $D$ as a function of diffusion power, with a linear fit. (c) Best-fit $\delta$-peak ionisation rate $S$ (Eq. \ref{fokker-planck}) for the different diffusion powers. The diffusion light was resonant with the $E_y$ transition.}
    \label{fig:optic_sup_diff}
\end{figure*}
When the optical frequency has a definite value $f$, the photon counts as a function of laser detuning $\Delta=f-f_\mathrm{laser}$ are given by the homogeneous line\begin{align}
    C_h(\Delta) = C_0 \frac{(\gamma_h/2)^2}{(\Delta^2 + (\gamma_h/2)^2)}.
\end{align}
Note that in this case, $\gamma_h$ may be power-broadened by the laser pulses.

The photon counts for a given $P(f,\tau_d)$ can then be found by convolution: \begin{align}\label{eq:sup_ou_cts}
    C(f,\tau_d) = \int_{-\infty}^\infty P(f',\tau_d)C_h(f-f')df'.
\end{align}
Here, we are interested in $C(0, \tau_d)$, because the laser pulses during the first and second check blocks $C_1$ and $C_2$ are applied at the same frequency (Fig. \ref{fig:optical}f). $C(f, \tau_d)$ could be probed by detuning $C_1$ and $C_2$, which could give more insight into the diffusion dynamics, though this is beyond the scope of the present work.

We initially used the model from van de Stolpe \textit{et al.} \cite{vandestolpeCheckprobeSpectroscopyLifetimelimited2025}. However, in our experiments the inhomogeneous linewidth $\gamma_i$ and the homogeneous linewidth $\gamma_h$ only differ by a factor $\sim5$, so the treatment of the optical frequency as a free random walker can only be valid for short diffusion times. This is because the probability distribution of the optical frequency should converge to the inhomogeneous distribution, while a free random walker results in an infinitely broad distribution as the diffusion time $\tau_d \rightarrow \infty$.

Instead, we use an Ornstein-Uhlenbeck (O-U) diffusion process to model $P(f,t)$. The main motivation is that, unlike the free random walk, $P(f,\tau_d)$ for an O-U process reproduces the gaussian inhomogeneous line as $\tau_d\rightarrow\infty$. In other recent work, a microscopic discrete-charges model is used to further motivate the use of an O-U model \cite{bownessLaserInducedSpectralDiffusion2025}. While the microscopic discrete nature of the charge-hopping dynamics can be complex \cite{pieplowQuantumElectrometerTimeresolved2025}, the O-U model works sufficiently well to model the observed photon counts.

In O-U diffusion, $P(f,t)$ is given by a gaussian with time-dependent mean $\mu(t)$ and variance $V(t)$ \begin{align}
    P(f, t) &= \frac{1}{\sqrt{2 \pi V(t)}} \exp{\left[-\frac{(f-\mu(t))^2}{2V(t)}\right]},\\
    \mu(t) &= f_i e^{-\theta t} + f_0(1-e^{-\theta t}),\\
    V(t) &= \frac{D}{\theta} (1-e^{-2\theta t}).
\end{align}
$f_0$ is the central frequency of the inhomogeneous distribution, that is, it is the frequency that the emitter will generally tend towards, and $f_i$ is the starting frequency. We ensure $f_0\approx f_i$ by setting the laser frequencies at the centre of the inhomogeneous line. The other two parameters of the O-U process are the diffusion parameter $D$, in units of $\mathrm{MHz}^2\mathrm{s}^{-1}$, and the parameter $\theta$ in units of $\mathrm{s}^{-1}$. To ease the physical interpretation we replace $\theta$ with the asymptotic linewidth of the inhomogeneous distribution $\mathrm{\gamma_i}$ as model parameter, which are related by \begin{align}
    \theta = D(2\sqrt{2\ln2} /\gamma_i)^2.
\end{align}

To calculate the photon counts $C(0, \tau_d)$ (Eq. \ref{eq:sup_ou_cts}), a value for $\gamma_h$ is required. While it is possible to include it as a fit parameter, this is not desirable as there exists a strong covariance between $\gamma_i$ and $\gamma_h$. Instead, we estimate the power-broadened $\gamma_h$ at the 5 nW power used for the $f_{RO}$ and $f_{SP}$ pulses during the check blocks using PLE data. While we do not have PLE data at 5 nW, we did measure a PLE at 2 nW, which had a FWHM of 17.3(4) MHz. The power-broadened $\gamma_h$ is given by \cite{hermans_entangling_2023}\begin{align}
    \gamma_{h}(P) = \sqrt{\gamma_0^2 + bP},
\end{align}
with $\gamma_0\approx13$ MHz the lifetime limited linewidth. An upper bound for the power-broadened line at 5 nW is then found by extrapolating from the linewidth at 2 nW, which results in $\gamma_h(5~ \mathrm{nW})\approx22$ MHz. If the 2~nW data is not power-broadened, then $\gamma_h$ at 5~nW, and the model-predicted $\gamma_{i}$, will be narrower. This leaves $\gamma_i$, $C_0$, and $D$ as free parameters. 

We measured the spectral diffusion at 9 different diffusive laser powers (Fig. \ref{fig:optic_sup_diff}). Because we expect that $\gamma_i$ is independent of the laser power, we fit one model jointly to all backward diffusion data, with a separate $C_0$ and $D$ for each laser power, but using a single $\gamma_i$. The motivation is that $\gamma_i$ should only be a function of the photon energy, as the photon energy determines the amount of charge traps that can be ionised or populated by the laser radiation, and the number of traps determines the total frequency range that can be explored by spectral diffusion. Even with this constraint the model fits the data rather well. This is additionally motivated by a reasonable reduced chi-square value of 2.1. 

\section{O-U diffusion with ionisation}
\label{sec:supp_ion}
In this section we derive a model that combines O-U diffusion of the emitter frequency with frequency-dependent ionisation. 

At low powers, there is no significant difference between the forward and backward-time correlations, apart from a $\sim$4\% decrease in photon counts due to ionisation during the check blocks. However, at higher powers, ionisation of the NV centre additionally reduces the photon counts in the forward-time correlations, because after a high photon-count rate in the first check block heralds that the emitter is on-resonance, the diffusion pulse can cause ionisation, resulting in no photons being emitted in the second check block. In this situation, calculating $P(f,t)$ becomes complex because there is a probability that the emitter disappears when its frequency is close to the laser frequency. To reach an approximate solution, we turn to the Fokker-Planck description of O-U diffusion, and treat ionisation as a $\delta$-function like probability sink at $f=0$. While a more realistic approach could take into account the spectral response to the excitation light, we analyse only the $\delta$-function approach here.

The evolution of $P(f,\tau_d)$ in the presence of a $\delta$ sink at $f_i$ is described by \begin{equation}
\label{fokker-planck}
    \frac{\partial P(f,\tau_d)}{\partial \tau_d} = \theta \frac{\partial}{\partial f} (f P(f,\tau_d)) + D \frac{\partial^2 P(f,\tau_d)}{\partial f^2} - S\delta(f-f_i) P(f,\tau_d),
\end{equation}
where $S$ in units if $\mathrm{s}^{-1}$ is the strength of the sink. To calculate a solution, we follow the approach set out by Chow and Liu \cite{chowFokkerPlanckEquation2004} who provide solutions for a general class of Fokker-Planck equations. By taking the Laplace transform of Eq. \ref{fokker-planck}, the Laplace transform of the solution to the O-U model with ionisation $\tilde{P}(f,s)$ can be expressed in terms of the Laplace transform of the problem without ionisation $\tilde{P}_0(f,s)$: \begin{align}\label{Pfs}
    \tilde{P}(f,s) = \tilde{P}_0(f,s)\left(1 - \frac{S\tilde{P}_0(0,s)}{1+S\tilde{P}_0(0,s)}\right).
\end{align}
For brevity, we have assumed that the ionising laser is at $f_i=0$. Again following Chow and Liu \cite{chowFokkerPlanckEquation2004}, to calculate $\tilde{P}_0(f,s)$, we use the eigenfunction expansion of the sinkless solution $P_0(f,\tau_d)$: \begin{align}
    P_0(f,\tau_d) = \sum_{n=0}^{\infty}\psi_0(f)\psi_n(f)\left[\frac{\psi_n(0)}{\psi_0(0)}\right] e^{\lambda_n \tau_d}=\sum_{n=0}^{\infty}w_n(f)e^{\lambda_n \tau_d}.
\end{align}
Our sought-after Laplace transform of the sinkless solution can now be expressed as \begin{align}\label{p0fs}
    \tilde{P}_0(f,s) = \sum_{n=0}^{\infty}w_n(f) (\lambda_n + s)^{-1}
\end{align}
The eigenfunctions $\psi_n(f)$ are solutions to the equation\begin{align}
    \left[ -\frac{1}{2} \frac{\mathrm{d}^2}{\mathrm{d}f^2} + \frac{\theta^2}{8D^2} - \frac{\theta}{4D}\right]\psi_n(f)=E_n\psi_n(f),
\end{align} with $E_n=\lambda_n/2D$.
This has a similar form as the standard quantum harmonic oscillator, with an offset in energy, and so the eigenfunctions and eigenvalues can easily be found: \begin{align}
    \lambda_n &= n\theta,\\
    \psi_n(f) &= \frac{1}{\sqrt{2^n\ n!}}\left(\frac{\theta}{2\pi D}\right)^{1/4}\exp{\left(-\frac{\theta f^2}{4D}\right)} H_n\left(f\sqrt{\theta/2D}\right),
\end{align}
where $H_n$ are Hermite polynomials. 

What remains is to calculate $\tilde{P}(f,s)$ and find the time-domain solution $P(f,\tau_d)$. For our model, we numerically calculate $\tilde{P}_0(f,s)$ using Eq. \ref{p0fs}, truncating to the first $N_{\mathrm{eigen}}=2000$ eigenfunctions. Note how the solution is only valid for $\tau_d \gg 1/(\theta N_{\mathrm{eigen}})$, as each eigenfunction only decays as $e^{-n\theta \tau_d}$. To avoid numerical overflow for large $n$, we use the identity \cite{AbramowitzStegun} $$e^{-x^2/2}H_n(x) \approx \frac{2^n}{\sqrt{\pi}}\Gamma\left(\frac{n+1}{2}\right)\cos(x\sqrt{2n}-n\pi/2)$$ to first calculate \begin{equation}
\exp{\left(-\frac{\theta f^2}{4D}\right)} \frac{H_n\left(f\sqrt{\theta/2D}\right) }{\sqrt{2^n\ n!}}\approx 
\exp\left[\frac{n \ln 2}{2} + \ln\Gamma\left[\frac{n+1}{2}\right]  - \frac{\ln\Gamma\left[n+1\right]}{2} \right] \cos\left[\frac{f\sqrt{n\theta}}{\sqrt{D}} - \frac{n\pi}{2} \right],
\end{equation}
directly calculating the $\ln \Gamma$ function using the Lanczos approximation \cite{lanczosPrecisionApproximationGamma1964}. To find $P(f,\tau_d)$, we calculate $\tilde{P(f,s)}$ using Eq. \ref{Pfs}, and use a numerical inverse Laplace transform \cite{abateMultiprecisionLaplaceTransform2004}. The photon counts in the diffusion experiment are again calculated by convolving the homogeneous Lorentzian lineshape with $P(f,\tau_d)$, so $P(f,\tau_d)$ needs to be calculated for a large enough range around the laser frequency.

As already stated, in all the diffusion data (Fig. \ref{fig:optic_sup_diff}) the forward correlation counts are $\sim4\%$ lower than the reverse counts, which we attribute to ionisation already occurring during the $C_1$ and $C_2$ pulses. After rescaling the forward counts by this factor, we fit the model that includes ionisation, discussed above, while only varying the ionisation rate $S$. The agreement of the fit to the data confirms that the Laplace-transform approach correctly calculates $P(f, \tau_d)$, and that the $\delta$-sink representation of frequency-dependent ionisation is able to reproduce our data.

We note that our analysis here does not explicitly take into account the possibility that the diffusion pulse, resonant with the $E_y$ ($m_s=0$) optical transition, can pump the electron-spin from the $m_s=0$ to the $m_s=\pm1$ spin ground states. This process would limit the ionisation probability, as the diffusion pulse cannot ionize the NV centre from the $m_s=\pm1$ ground states. The magnitude of this effect depends on the probability to decay to the $m_s=\pm1$ states from $E_y$, compared to the probability to ionize from the $E_y$ excited state. 

\section{Spectral diffusion comparison}
\label{sec:qne}

We measured the spectral diffusion of an NV centre that was previously used for remote entanglement experiments \cite{stolkMetropolitanscaleHeraldedEntanglement2024} to compare the spectral diffusion seen in our diamond (Fig. \ref{fig:reference}). This NV centre comes from a (100)-grown, natural abundance diamond (element Six). Diffusion was measured at diffusion powers $P_d$ up to 5 \textmu W, however, higher powers caused almost complete ionisation of the NV centre, making it impossible to extract the spectral diffusion from the reverse photon correlations. Therefore only the measurements at 250 nW and 500 nW were used. For the check blocks laser pulses at $f_{RO}$ and $f_{SP}$ of 0.8 nW for 250 us were used. Check-probe PLE measurements were performed yielding $\gamma_h=22$ MHz at this power.

The resulting fit parameters are displayed in Table \ref{tab:OU}. The difference in diffusion parameter $D$ implies that the diffusion timescales are $\sim30$ times faster in our sample. This is also visible in the data; the photon counts drop a factor $\sim30$ earlier in our sample. The slower spectral diffusion also explains the observed increase in ionisation, which is only possible when the diffusion laser is resonant with the NV centre's optical transition.
\begin{figure}[h]
    \includegraphics{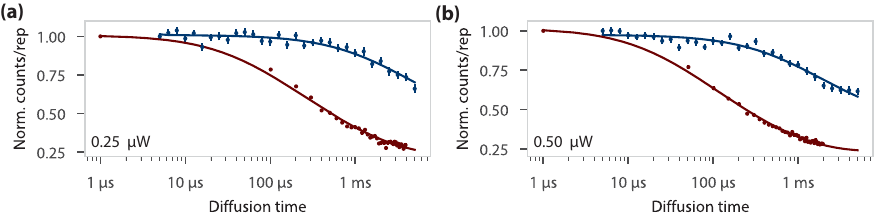}
    \caption{\emph{Spectral diffusion in another sample.} Diffusion measurements of NV centres in Fukuoka (red, grown in the present work) and an NV in a natural abundance (100)-grown diamond, previously used for remote entanglement experiments (blue) \cite{stolkMetropolitanscaleHeraldedEntanglement2024}, at (a) $P_d=250$ nW and (b) $P_d=500$ nW. Solid lines are fits by the O-U model.}
    \label{fig:reference}
\end{figure}
\begin{table}[h]
    \centering
    \begin{tabular}{llll}\toprule
         Sample&   $\gamma_{i,\mathrm{red}}$ &$D(250 \mathrm{nW)}$ &$D(500 \mathrm{nW)}$\\\midrule
         Fukuoka&  117(2)~MHz &0.34(4)~MHz$^2$/nW &0.75(7)~MHz$^2$/nW \\
 From  \cite{stolkMetropolitanscaleHeraldedEntanglement2024}& 42(9)~MHz &0.011(1)~MHz$^2$/nW &0.025(2)~MHz$^2$/nW \\ \end{tabular}
    \caption{\emph{Spectral diffusion model parameters} 
    }
    \label{tab:OU}
\end{table}

\newpage

\end{widetext}

\end{document}